\documentclass[preprint,noshowpacs,noshowkeys,aps]{revtex4}%
\usepackage{amsfonts}
\usepackage{amsmath}
\usepackage{amssymb}
\usepackage{graphicx}%
\providecommand{\U}[1]{\protect\rule{.1in}{.1in}}

\begin{document}
\preprint{ }
\title[Nonlocal Phases of Wavefunctions]{Beyond the Dirac phase factor$\boldsymbol{:}$ Dynamical Quantum
Phase-Nonlocalities in the Schr\"{o}dinger Picture}
\author{Konstantinos Moulopoulos}
\email{cos@ucy.ac.cy}
\affiliation{University of Cyprus, Department of Physics, 1678 Nicosia, Cyprus}

\begin{abstract}
Generalized solutions of the standard gauge transformation equations are
presented and discussed in physical terms. They go beyond the usual Dirac
phase factors and they exhibit nonlocal quantal behavior, with the well-known
Relativistic Causality of classical fields affecting directly the
\textit{phases} of wavefunctions in the Schr\"{o}dinger Picture. These
nonlocal phase behaviors, apparently overlooked in path-integral approaches,
give a natural account of the dynamical nonlocality character of the various
(even static) Aharonov-Bohm phenomena, while at the same time they seem to
respect Causality. For particles passing through nonvanishing magnetic or
electric fields they lead to cancellations of Aharonov-Bohm phases at the
observation point, generalizing earlier semiclassical experimental
observations (of Werner \& Brill) to delocalized (spread-out) quantum states.
This leads to a correction of previously unnoticed sign-errors in the
literature, and to a natural explanation of the deeper reason why certain
time-dependent semiclassical arguments are consistent with static results in
purely quantal Aharonov-Bohm configurations. These nonlocalities also provide
a remedy for misleading results propagating in the literature (concerning an
uncritical use of Dirac phase factors, that persists since the time of
Feynman's work on path integrals). They are shown to conspire in such a way as
to exactly cancel the instantaneous Aharonov-Bohm phase and recover
Relativistic Causality in earlier \textquotedblleft
paradoxes\textquotedblright\ (such as the van Kampen
thought-experiment),$\boldsymbol{\ }$and to also complete Peshkin's discussion
of the electric Aharonov-Bohm effect in a \textit{causal} manner. The present
formulation offers a direct way to address time-dependent single- \textit{vs}
double-slit experiments and the associated causal issues -- issues that have
recently attracted attention, with respect to the inability of current
theories to address them.

KEYWORDS$\boldsymbol{:}$ Aharonov-Bohm, Gauge Transformations, Dirac Phase
Factor, Quantum Phases

\end{abstract}
\volumeyear{2011}
\volumenumber{number}
\issuenumber{number}
\eid{identifier}
\date[August 30, 2011]{}
\maketitle

\section{Introduction}

\bigskip The Dirac phase factor $-$ with a phase containing spatial or
temporal integrals of potentials (of the general form $%
{\displaystyle\int\limits^{\boldsymbol{r}}}
\boldsymbol{A}\cdot d\boldsymbol{x}^{\prime}-c%
{\displaystyle\int\limits^{t}}
\phi dt^{\prime}$) $-$ is the standard and widely used solution of the gauge
transformation equations of Electrodynamics (with $\boldsymbol{A}$ and $\phi$
vector and scalar potentials respectively). In a quantum mechanical context,
it connects wavefunctions of two systems (with different potentials) that
experience the same classical fields at the observation point $(\boldsymbol{r}%
,t)$, the two more frequently discussed cases being: either systems that are
completely gauge-equivalent (a trivial case with no physical consequences), or
systems that exhibit phenomena of the Aharonov-Bohm type (magnetic or
electric)\cite{AB} $-$ and then this Dirac phase has nontrivial observable
consequences (mathematically, this being due to the fact that the
corresponding \textquotedblleft gauge function\textquotedblright\ is now
multiple-valued). In the above two cases, the classical fields experienced by
the two (mapped)\ systems are equal \textit{at every point }of the accessible
spacetime region. However, it has not been widely realized that the gauge
transformation equations, viewed in a more general context, can have
\textit{more general solutions} than simple Dirac phases, and these lead to
wavefunction-\textit{phase-nonlocalities} that have been widely overlooked and
that seem to have important physical consequences. These nonlocal solutions
are applicable to cases where the two systems are allowed to experience
\textit{different fields} at spacetime points (or regions) that are
\textit{remote} to (and do \textit{not} contain) the observation point
$(\boldsymbol{r},t)$ (these regions being \textit{physically accessible} to
the particle, unlike genuine Aharonov-Bohm cases). In this article we
rigorously show the existence of these generalized solutions, demonstrate them
in simple physical examples, and fully explore their structure, presenting
cases (and closed analytical results for the wavefunction-phases) that
actually connect (or map) two quantal systems that are \textbf{neither
physically equivalent nor of the usual Aharonov-Bohm type}. We also fully
investigate the consequences of these generalized (\textit{nonlocal})
influences (on wavefunction-phases) and find them to be numerous and
important; we actually find them to be of a different type in static and in
time-dependent field-configurations (and in the latter cases we show that they
lead to Relativistically \textit{causal} behaviors, that apparently resolve
earlier \textquotedblleft paradoxes\textquotedblright\ arising in the
literature from the use of standard Dirac phase factors). The nonlocal phase
behaviors discussed in the present work may be viewed as a justification for
the (recently emphasized\cite{Popescu}) terminology of \textquotedblleft
dynamical nonlocalities\textquotedblright\ associated with all Aharonov-Bohm
effects (even static ones), although in our approach these nonlocalities seem
to also respect Causality (without the need to independently invoke the
Uncertainty Principle) -- and, to the best of our knowledge, this is the first
theoretical picture with such characteristics.

In order to introduce some background and further motivation for this article
let us first remind the reader of a very basic property that will be central
to everything that follows, which however is usually taken to be valid only in
a restricted context (but is actually more general than often realized). This
property is a simple (U(1)) phase-mapping between quantum systems, and is
usually taken in the context of gauge transformations, ordinary or
singular$\boldsymbol{;}$ here, however, it will appear in a more general
framework, hence the importance of reminding of its independent, basic and
more general origin. We begin by recalling that, if $\Psi_{1}(\mathbf{r},t)$
and $\Psi_{2}(\mathbf{r},t)$ are solutions of the time-dependent
Schr\"{o}dinger (or Dirac) equation for a quantum particle of charge $q$ that
moves (as a test particle) in two distinct sets of (predetermined and
classical) vector and scalar potentials ($\mathbf{A}_{1},\phi_{1}$) and
($\mathbf{A}_{2},\phi_{2}$), that are generally spatially- and
temporally-dependent [and such that, at the spacetime point of observation
$(\mathbf{r},t)$, the magnetic and electric fields are the same in the two
systems], then we have the following formal connection between the solutions
(wavefunctions) of the two systems%

\begin{equation}
\Psi_{2}(\mathbf{r},t)=e^{i\frac{q}{\hbar c}\Lambda(\mathbf{r},t)}\Psi
_{1}(\mathbf{r},t), \label{Basic1}%
\end{equation}
with the function $\Lambda(\mathbf{r},t)$ required to satisfy%

\begin{equation}
\nabla\Lambda(\mathbf{r,t})=\mathbf{A}_{2}(\mathbf{r},t)-\mathbf{A}%
_{1}(\mathbf{r},t)\qquad and\qquad-\frac{1}{c}\frac{\partial\Lambda
(\mathbf{r},t)}{\partial t}=\phi_{2}\left(  \mathbf{r},t\right)  -\phi
_{1}(\mathbf{r},t). \label{Basic11}%
\end{equation}
The above property can be immediately proven by substituting each $\Psi_{i}$
into its corresponding ($i_{th}$) time-dependent Schr\"{o}dinger equation
(namely with the set of potentials ($\mathbf{A}_{i}(\mathbf{r},t),\phi
_{i}(\mathbf{r},t)$))$\boldsymbol{:}$ one can then easily see that
(\ref{Basic1}) and (\ref{Basic11}) guarantee that both Schr\"{o}dinger
equations are indeed satisfied together (after cancellation of a few terms and
then elimination of a global phase factor in system 2). [In addition, the
equality of all classical fields at the observation point, namely
$\mathbf{B}_{2}(\mathbf{r},t)=\nabla\times\mathbf{A}_{2}(\mathbf{r}%
,t)=\nabla\times\mathbf{A}_{1}(\mathbf{r},t)=$ $\mathbf{B}_{1}(\mathbf{r},t)$
for the magnetic fields and $\mathbf{E}_{2}(\mathbf{r},t)=-\nabla\phi
_{2}\left(  \mathbf{r},t\right)  -\frac{1}{c}\frac{\partial\mathbf{A}%
_{2}(\mathbf{r},t)}{\partial t}=-\nabla\phi_{1}\left(  \mathbf{r},t\right)
-\frac{1}{c}\frac{\partial\mathbf{A}_{1}(\mathbf{r},t)}{\partial t}%
=\mathbf{E}_{1}(\mathbf{r},t)$ for the electric fields, is obviously
consistent with all equations (\ref{Basic11}) (as is easy to see if we take
the \textit{curl} of the 1st and the \textit{grad} of the 2nd) $-$ provided,
at least, that $\Lambda(\mathbf{r},t)$ is such that interchanges of partial
derivatives with respect to all spatial and temporal variables (at the point
$\left(  \mathbf{r},t\right)  $) are allowed].

As already mentioned, the above fact is of course well-known within the
framework of the theory of quantum mechanical gauge transformations (the usual
case being for $\mathbf{A}_{1}=\phi_{1}=0$, hence for a mapping from a system
with no potentials)$\boldsymbol{;}$ but in that framework, these
transformations are supposed to connect (or map) two \textit{physically
equivalent systems} (more rigorously, this being true for ordinary gauge
transformations, in which case the function $\Lambda(\mathbf{r},t)$, the
so-called gauge function, is unique (single-valued) in spacetime coordinates).
In a formally similar manner, the above argument is also often used in the
context of the so-called \textquotedblleft singular gauge
transformations\textquotedblright, where $\Lambda$ is multiple-valued, but the
above equality of classical fields is still imposed (at the observation point,
which always lies in a physically accessible region)$\boldsymbol{;}$ then the
above simple phase mapping (at all points of the physically accessible
spacetime region, that always and everywhere experience equal fields) leads to
the standard phenomena of the Aharonov-Bohm type, where \textit{unequal fields
in physically inaccessible regions} have observable consequences. However, we
should keep in mind that that above property ((\ref{Basic1}) and
(\ref{Basic11}) taken together) can be \textit{more generally valid} $-$ and
in this article we will present cases (and closed analytical results for the
appropriate phase connection $\Lambda(\mathbf{r},t)$) that actually connect
(or map) two systems (in the sense of (\ref{Basic1})) that are \textit{neither
physically equivalent nor of the usual Aharonov-Bohm type}. And naturally,
because of the above provision of field equalities at the observation point,
it will turn out that any nonequivalence of the two systems will involve
\textit{remote} (although \textit{physically accessible}) regions of
spacetime, namely regions that do \textit{not} contain the observation point
$(\mathbf{r},t)$ (and in which regions, as we shall see, the classical fields
experienced by the particle may be \textit{different} in the two systems).

\section{Motivation}

One may wonder on the actual reasons why one should be looking for more
general cases of a simple phase mapping of the type (\ref{Basic1}) between
\textit{nonequivalent} systems. To answer this, let us take a step back and
first recall some simple and well-known results that originate from the above
phase mapping. It is standard knowledge, for example, that, if we want to find
solutions $\Psi(x,t)$ of the $t$-dependent Schr\"{o}dinger (or Dirac) equation
for a quantum particle (of charge $q)$ that moves along a (generally curved)
one-dimensional (1-D) path, and in the presence (somewhere in the embedding
3-dimensional (3-D) space) of a fairly localized (and time-independent)
classical magnetic flux $\Phi$ that \textit{does \textbf{not} pass through any
point of the path,} then we formally have%

\begin{equation}
\Psi(x,t)^{(\mathbf{A})}\sim e^{i\frac{q}{\hbar c}\int_{x_{0}}^{x}%
\mathbf{A(r}^{\prime})\cdot d\mathbf{r}^{\prime}}\Psi(x,t)^{(0)}
\label{standard}%
\end{equation}
(the dummy variable $\mathbf{r}^{\prime}$ describing points along the 1-D
path, and the term \textquotedblleft formally\textquotedblright\ signifying
that the above is valid \textit{before imposition of any boundary conditions
}(meaning that these are to be imposed only on $\Psi^{(\mathbf{A})}$ and not
necessarily on $\Psi^{(\mathbf{0})}$). In (\ref{standard}), $\Psi(x,t)^{(0)}$
is a formal solution of the same system in the case of absence of any
potentials (hence with magnetic flux $\Phi=0$ everywhere in the 3-D space).
The above holds because, for \textit{all points} $\mathbf{r}^{\prime}$ of the
1-D path, the particle experiences a vector potential $\mathbf{A(r\prime)}$ of
the form $\mathbf{A(r\prime)=\nabla}^{\prime}\Lambda\mathbf{(r}^{\prime
}\mathbf{)}$ (since the magnetic field is $\mathbf{\nabla}^{\prime
}\mathbf{\times A(r}^{\prime})=0$\textbf{ }for \textit{all} $\mathbf{r}%
^{\prime}$, by assumption), in combination with the above phase-mapping (with
a phase $\frac{q}{\hbar c}\Lambda\mathbf{(r)}$) between two quantum systems,
one in the presence and one in the absence of a vector potential (i.e. the
potentials in (\ref{Basic11}) being $\mathbf{A}_{1}=0$ and $\mathbf{A}_{2}=$
$\mathbf{A}$, together with $\phi_{2}=\phi_{1}=0$ if we decide to attribute
everything to vector potentials only). In this particular system, the obvious
$\Lambda\mathbf{(r)}$ that solves the above $\mathbf{A(\mathbf{r})=\nabla
}\Lambda\mathbf{(r)}$ (for \textbf{all} points of the 1-D space available to
the particle) is indeed $\Lambda\mathbf{(r)=}\Lambda\mathbf{(r}_{0}%
\mathbf{)+}\int_{\mathbf{r}_{0}}^{\mathbf{r}}\mathbf{A(r}^{\prime})\cdot
d\mathbf{r}^{\prime}$ (since $\int_{1}^{2}\mathbf{\nabla}^{\prime}%
\Lambda\mathbf{(r}^{\prime}\mathbf{)}\cdot d\mathbf{r}^{\prime}=\Lambda
\mathbf{(}2\mathbf{)-}\Lambda\mathbf{(}1\mathbf{)}$), and this gives
(\ref{standard}) (if $\mathbf{r}$ denotes the above point $x$ of observation
and $\mathbf{r}_{0}$ the arbitrary initial point $x_{0}$ (both lying on the
physical path), and if the constant $\mathbf{\Lambda(r}_{0}\mathbf{)}$ is
taken to be zero).

What if, however, some parts of the magnetic field that comprise the magnetic
flux $\Phi$ actually \textit{pass through} some points or a whole region
(interval) of the path available to the particle? In such a case, the above
standard argument is not valid (as $\mathbf{A}$ cannot be written as a grad at
any point of the interval where the magnetic field $\mathbf{\nabla\times
A}\neq0$). Are there however general results that we can still write for
$\Psi(x,t)^{(\mathbf{A})}$, if the spatial point of observation $x$ is again
outside the interval with the nonvanishing magnetic field? Or, what if in the
previous problems, the magnetic flux (either remote, or partly passing through
the path) is time-dependent $\Phi(t)$? (In that case then, there exists in
general an additional electric field $E$ induced by Faraday's law of Induction
on points of the path, and the usual gauge transformation argument is once
again not valid).

Returning to another standard (solvable) case (which is actually the
\textquotedblleft dual\textquotedblright\ or the \textquotedblleft electric
analog\textquotedblright\ of the above), if along the 1-D physical path the
particle experiences only a spatially-uniform (but generally time-dependent)
classical scalar potential $\phi(t)$, we can again formally map $\Psi
(x,t)^{(\phi)}$ to a potential-free solution $\Psi(x,t)^{(0)}$, through a
$\Lambda(t)$ that now solves $-\frac{1}{c}\frac{\partial\Lambda(t)}{\partial
t}=\phi(t)$, and this gives $\Lambda(t)=\Lambda(t_{0})-c\int_{t_{0}}^{t}%
\phi(t^{\prime})dt^{\prime}$, leading to the \textquotedblleft electric
analog\textquotedblright\ of (\ref{standard}), namely%

\begin{equation}
\Psi(x,t)^{(\phi)}\sim e^{-i\frac{q}{\hbar}\int_{t_{0}}^{t}\phi(t^{\prime
})dt^{\prime}}\Psi(x,t)^{(0)} \label{standard2}%
\end{equation}
with obvious notation. (Notice that, for either of the two mapped systems in
this problem, the electric field is zero at all points of the path). What if,
however, the scalar potential has also some $x$-dependence along the path
(that leads to an electric field (in a certain interval) that the particle
passes through)? In such a case, the above standard argument is again not
valid. Are there however general results that we can still write for
$\Psi(x,t)^{(\phi)}$, if the spatial point of observation $x$ is again outside
the interval with the nonvanishing electric field?

We state here directly that this article will provide affirmative answers to
questions of the type posed above, by actually giving the corresponding
general results in closed analytical forms.

At this point it is also useful to briefly reconsider the earlier mentioned
case, namely of a time-dependent $\Phi(t)$ that is remote to the 1-D physical
path, because in this manner we can immediately provide another motivation for
the present work$\boldsymbol{:}$ this time-dependent problem is surrounded
with a number of important misconceptions in the literature (the same being
true about its electric analog, as we shall see)$\boldsymbol{:}$ the formal
solution that is usually written down for a $\Phi(t)$ is again (\ref{standard}%
), namely the above spatial line integral of $\mathbf{A}$, in spite of the
fact that $\mathbf{A}$ is now $t$-dependent$\boldsymbol{;}$ the problem then
is that, because of the first of (\ref{Basic11}), $\Lambda$ must now have a
$t$-dependence and, from the second of (\ref{Basic11}), there must necessarily
be scalar potentials involved in the problem (which have been by force set to
zero, in our pre-determined mapping between vector potentials only). Having
decided to use systems that experience only vector (and not scalar)
potentials, the correct solution cannot be simply a trivial $t$-dependent
extension of (\ref{standard}). A corresponding error is usually made in the
electric dual of the above, namely in cases that involve $\mathbf{r}%
$-dependent scalar potentials, where (\ref{standard2}) is still erroneously
used (with $\phi(t^{\prime})$ replaced by $\phi(\mathbf{r,}t^{\prime})$),
giving an $\mathbf{r}$-dependent $\Lambda$, although this would necessarily
lead to the involvement of vector potentials (through the first of
(\ref{Basic11}) and the $\mathbf{r}$-dependence of $\Lambda$) that have been
neglected from the beginning $-$ a situation (and an error) that appears, in
exactly this form, in the description of the so-called electric Aharonov-Bohm
effect\cite{AB,Peshkin} as we shall see.

Speaking of errors in the literature, it might here be the perfect place to
also point to the reader the most common misleading statement often made in
the literature (and again, for notational simplicity, we restrict our
attention to a one-dimensional system, with spatial variable $x$, although the
statement is obviously generalizable to (and often made for systems of) higher
dimensionality by properly using line integrals over arbitrary curves in
space)$\boldsymbol{:}$ It is usually stated [e.g. in Brown \&
Holland\cite{BrownHolland}, see i.e. their eq. (57) applied for vanishing
boost velocity $\mathbf{v}=0$] that the general gauge function that connects
(through a phase factor $e^{i\frac{q}{\hbar c}\Lambda(x,t)}$) the
wavefunctions of a quantum system with no potentials (i.e. with a set of
potentials $(\mathbf{0},0)$) to the wavefunctions of a quantum system that
moves in vector potential $\boldsymbol{A}(x,t)$ and scalar potential
$\phi(x,t)$ (i.e. in a set of potentials $(\boldsymbol{A},\phi)$) is the
obvious combination (and a natural extension) of (\ref{standard}) and
(\ref{standard2}), namely%

\begin{equation}
\Lambda(x,t)=\Lambda(x_{0},t_{0})+%
{\displaystyle\int\limits_{x_{0}}^{x}}
\boldsymbol{A}(x^{\prime},t)dx^{\prime}-c%
{\displaystyle\int\limits_{t_{0}}^{t}}
\phi(x,t^{\prime})dt^{\prime}, \label{wrong}%
\end{equation}
which, however, is \textbf{incorrect } for $x$ and $t$ uncorrelated
variables$\boldsymbol{:}$ it does \textbf{not }satisfy the standard
system\textbf{\ }of gauge transformation equations%

\begin{equation}
\nabla\Lambda(x,t)=\mathbf{A(}x,t)\mathbf{\qquad and\qquad-}\frac{1}{c}%
\frac{\partial\Lambda(x,t)}{\partial t}=\phi(x,t). \label{BasicPDE}%
\end{equation}
The reader can easily see why: (i) when the $\nabla$ operator acts on
eq.(\ref{wrong}), it gives the correct $A(x,t)$ from the 1st term, but it also
gives some annoying additional nonzero quantity from the 2nd term (that
survives because of the $x$-dependence of $\phi$); hence it invalidates the
first of the basic system (\ref{BasicPDE}). (ii) Similarly, when the
$\mathbf{-}\frac{1}{c}\frac{\partial}{\partial t}$ operator acts on
eq.(\ref{wrong}), it gives the correct $\phi(x,t)$ from the 2nd term, but it
also gives some annoying additional nonzero quantity from the 1st term (that
survives because of the $t$-dependence of $\mathbf{A}$); hence it invalidates
the second of the basic system (\ref{BasicPDE}). It is only when $\mathbf{A}$
is $t$-independent, and $\phi$ is spatially-independent, that eq.(\ref{wrong})
can be correct (as the above annoying terms do not appear and the basic system
is satisfied), although it is still not necessarily the most general form for
$\Lambda$, as we shall see. [An alternative form that is also widely thought
to be correct is again eq.(\ref{wrong}), but with the variables that are not
integrated over implicitly assumed to belong to the initial point (hence a
$t_{0}$ replaces $t$ in $\mathbf{A}$, and simultaneously an $x_{0}$ replaces
$x$ in $\phi$). However, one can see again that the system (\ref{BasicPDE}) is
not satisfied (the above differential operators, when acted on $\Lambda$, give
$\mathbf{A}(x,t_{0})$ and $\phi(x_{0},t)$, hence not the values of the
potentials at the point of observation $(x,t)$ as they should), this not being
an acceptable solution either].

What is the problem here? Or, better put, what is the deeper reason for the
above inconsistencies? The short answer is the uncritical use of Dirac phase
factors that come from path-integral treatments. It is indeed obvious that the
form (\ref{wrong}) that is often used in the literature (in canonical
(non-path-integral) formulations where $x$ and $t$ are \textbf{uncorrelated}
variables (and not correlated to produce a path $x(t)$)) \textit{is not
generally correct}, and that is one of the main points that has motivated this
work. We will find \textit{generalized results} that actually \textit{correct}
eq.(\ref{wrong}) through extra nonlocal terms, and through the proper
appearance of $x_{0}$ and $t_{0}$ (as in eq.(\ref{LambdaStatic1}) and
eq.(\ref{LambdaStatic2}) to be found later in Section III), and these are the
\textbf{exact} ones (namely the exact $\Lambda(x,t),$ that at the end, upon
action of $\nabla$ and $\mathbf{-}\frac{1}{c}\frac{\partial}{\partial t}$
satisfies exactly the basic system (\ref{BasicPDE}), viewed as a system of
Partial Differential Equations (PDEs)). And the formulation that gives these
results is generalized later in the article, for $\Lambda(x,y)$ (in the 2-D
static case) and also for $\Lambda(x,y,t)$ (in the full dynamical 2-D case),
and leads to the exact (nontrivial) forms of the phase function $\Lambda$ that
satisfy (in all cases) the system (\ref{BasicPDE}) $-$ with \textit{the direct
verification\ }(\textit{i.e. proof, by \textquotedblleft going
backwards\textquotedblright, that these forms are indeed the exact solutions
of }(\ref{BasicPDE}))\textit{ also being given }for the reader's
convenience\textit{. }[For the \textquotedblleft direct\textquotedblright\ and
rigorous mathematical derivations see \cite{jphysa}.]

This article gives a full exploration of issues related to the above
motivating discussion, by pointing to a \textquotedblleft
practical\textquotedblright\ (and generalized) use of gauge transformation
mapping techniques, that at the end lead to these generalized (and, at first
sight, unexpected) solutions for the general form of $\Lambda$. For cases such
as the ones discussed above, or even more involved ones, there still appears
to exist a simple phase mapping (between two inequivalent systems), but the
phase connection $\Lambda$ seems to contain not only integrals of potentials,
but also \textquotedblleft fluxes\textquotedblright\ of the classical
field-differences from \textit{remote} spacetime regions (regions, however,
that are physically accessible to the particle). The above mentioned systems
are the simplest ones where these new results can be applied, but apart from
this, the present investigation seems to lead to a number of nontrivial
corrections of misleading (or even incorrect) reports of the above type in the
literature, that are not at all marginal (and are due to an incorrect use of a
path-integral viewpoint in an otherwise canonical framework -- an error that
appears to have been made repeatedly since the original conception of the
path-integral formulation, as we shall see). The generalized $\Lambda$-forms
also lead to an honest resolution of earlier \textquotedblleft
paradoxes\textquotedblright\ (involving Relativistic Causality), and in some
cases to a new interpretation of known semiclassical experimental
observations, corrections of certain sign-errors in the literature, and
nontrivial extensions of earlier semiclassical results to general (even
completely delocalized) states. [As a byproduct, we will also show that --
contrary to what is usually stated in earlier but also recent popular reports
-- the semiclassical phase picked up by classical trajectories (that are
deflected by the Lorentz force) is \textit{opposite} (and not equal) to the
so-called Aharonov-Bohm phase due to the flux enclosed by the same
trajectories$\boldsymbol{;}$ we will also provide 2 figures to visually assist
the detailed proof of this result as well as to facilitate an elementary
physical understanding of this opposite sign relation]. Most importantly,
however, the new formulation seems capable of addressing causal issues in
time-dependent single- \textit{vs} double-slit experiments, an area that seems
to have recently attracted attention\cite{Tollaksen,Popescu,He}).

\section{1-d dynamic case}

\bigskip Let us begin with the simplest case of 1-D systems but in the most
general dynamic environment, i.e. a single quantum particle of charge $\ q$,
but in the presence of arbitrary (spatially nonuniform and time-dependent)
vector and scalar potentials. Let us actually consider this particle moving
either inside a set of potentials $A_{1}(x,t)$ and $\phi_{1}(x,t)$ (case 1) or
inside a set of potentials $A_{2}(x,t)$ and $\phi_{2}(x,t)$ (case 2), and try
to determine the most general gauge function $\Lambda(x,t)$ that takes us from
(maps) the wavefunctions of the particle in case 1 to those of the same
particle in case 2 (meaning the usual mapping (\ref{Basic1}) between the
wavefunctions of the two systems through the phase factor $\frac{q}{\hbar
c}\Lambda(x,t)$). [As already noted, we should keep in mind that for this
mapping to be possible we \textit{must} assume that at the point $(x,t)$ of
observation (or \textquotedblleft measurement\textquotedblright\ of $\Lambda$
or the wavefunction $\Psi$) \ we have equal electric fields ($E_{i}%
=-\nabla\phi_{i}-\frac{1}{c}\frac{\partial A_{i}}{\partial t}$), namely%

\begin{equation}
-\frac{\partial\phi_{2}(x,t)}{\partial x}-\frac{1}{c}\frac{\partial
A_{2}(x,t)}{\partial t}=-\frac{\partial\phi_{1}(x,t)}{\partial x}-\frac{1}%
{c}\frac{\partial A_{1}(x,t)}{\partial t} \label{Efield}%
\end{equation}
(so that the $A$'s and $\phi$'s in (\ref{Efield}) can indeed satisfy the basic
system of equations (\ref{Basic11}), or equivalently, of the system of
equations (\ref{xt-BasicSystem}) below $-$ as can be seen by taking the
$\frac{1}{c}\frac{\partial}{\partial t}$ of the 1st and the $\frac{\partial
}{\partial x}$ of the 2nd of the system (\ref{xt-BasicSystem}) and adding them
together)$\boldsymbol{.}$ But again, we will \textit{not} exclude the
possibility of the two systems passing through \textit{different }electric
fields in other regions of spacetime (that do \textit{not} contain the
observation point), i.e. for $(x^{\prime},t^{\prime})\neq(x,t)$. In fact, this
possibility \textbf{will come out naturally} from a careful solution of the
basic system (\ref{xt-BasicSystem})$\boldsymbol{;}$ it is for example
straightforward for the reader to immediately verify that the results
(\ref{LambdaStatic1}) or (\ref{LambdaStatic2}) that will be derived below (and
will contain contributions of electric field-differences from remote regions
of spacetime) indeed satisfy the basic input system of equations
(\ref{xt-BasicSystem}), something that will be explicitly verified below].

Returning to the question on the appropriate $\Lambda$ that takes us from the
set $(A_{1},\phi_{1})$ to the set $(A_{2},\phi_{2})$, we note again that, in
cases of static vector potentials ($A(x)$'s) \textit{and} spatially uniform
scalar potentials ($\phi(t)$'s) the form usually given for $\Lambda$ is the well-known%

\begin{equation}
\Lambda(x,t)=\Lambda(x_{0},t_{0})+\int_{x_{0}}^{x}A(x^{\prime})dx^{\prime
}-c\int_{t_{0}}^{t}\phi(t^{\prime})dt^{\prime} \label{LambdaUsual}%
\end{equation}
with $\ A(x)=A_{2}(x)-A_{1}(x)$ \ and $\ \phi(t)=\phi_{2}(t)-\phi_{1}(t)$ (and
it can be viewed as a combination of (\ref{standard}) and (\ref{standard2}),
being immediately applicable to the description of cases of \textit{combined}
magnetic and electric Aharonov-Bohm effects).

\bigskip But as already noted, even in the most general case, with
$t$-dependent $A$'s and $x$-dependent $\phi$'s (and with the variables$\ x$%
\ and$\ t$ being \textbf{completely uncorrelated}), it is often stated in the
literature that the appropriate $\Lambda$ has a form that is a plausible
extention of (\ref{LambdaUsual}), namely%

\begin{equation}
\Lambda(x,t)=\Lambda(x_{0},t_{0})+%
{\displaystyle\int\limits_{x_{0}}^{x}}
\left[  A_{2}(x^{\prime},t)-A_{1}(x^{\prime},t)\right]  dx^{\prime}-c%
{\displaystyle\int\limits_{t_{0}}^{t}}
\left[  \phi_{2}(x,t^{\prime})-\phi_{1}(x,t^{\prime})\right]  dt^{\prime},
\label{BrownHolland}%
\end{equation}
[with eq. (57) of Ref.\cite{BrownHolland}, taken for $\mathbf{v}=0$, being a
very good example to point to, since that article does not use a path-integral
language, but a canonical formulation with uncorrelated variables]. And as
already pointed out in Section II, this form is certainly \ \textit{incorrect}
\ for uncorrelated variables $x$ and $t$ \ (the reader can easily verify that
the system of equations (\ref{xt-BasicSystem}) below is ${\large not}$
satisfied by (\ref{BrownHolland}) $-$ see again Section II, especially the
paragraph after eq.(\ref{BasicPDE}))$\boldsymbol{.}$ We will find in the
present work that the correct form consists of two major modifications of
(\ref{BrownHolland})$\boldsymbol{:}$ (i) The first leads to the natural
appearance of a \textit{path} that continuously connects initial and final
points in spacetime, a property that (\ref{BrownHolland}) \textit{does not
have} [indeed, if the integration curves of (\ref{BrownHolland}) are drawn in
the $(x,t)$-plane, they do \textit{not} form a continuous path from
$(x_{0},t_{0})$ to $(x,t)$]. (ii) And the second modification is highly
nontrivial$\boldsymbol{:}$ it consists of nonlocal contributions of classical
electric field-differences from remote regions of spacetime. We will discuss
below the consequences of these terms and we will later show that such
nonlocal contributions also appear (in an extended form) in more general
situations, i.e. they are also present in higher spatial dimensionality (and
they then also involve remote magnetic fields in combination with the electric
ones)$\boldsymbol{;}$ these lead to modifications of ordinary Aharonov-Bohm
behaviors or have other important consequences, one of them being a natural
remedy of Causality \textquotedblleft paradoxes\textquotedblright\ in
time-dependent Aharonov-Bohm experiments.

The form (\ref{BrownHolland}) commonly used is of course motivated by the
well-known Wu \& Yang\cite{WuYang} nonintegrable phase factor, that has a
phase equal to \ $\int A_{\mu}dx^{\mu}=\int Adx-c\int\phi dt$, \ a form that
appears naturally within\ the framework of path-integral treatments, or
generally in physical situations where narrow wavepackets are implicitly
assumed for the quantum particle\textbf{:} the integrals appearing in
(\ref{BrownHolland}) are then taken along particle trajectories (hence spatial
and temporal variables \textit{not} being uncorrelated, but being connected in
a particular manner $x(t)$ to produce the path$\boldsymbol{;}$ all integrals
are therefore basically only time-integrals). But even then,
eq.(\ref{BrownHolland}) is valid only when these trajectories are always (in
time) and everywhere (in space) inside identical classical fields for the two
(mapped) systems. Here, however, we will be focusing on what a canonical (and
not a path-integral or other semiclassical) treatment leads to$\boldsymbol{;}$
this will cover the general case of arbitrary wavefunctions that can even be
completely spread-out in space, and will also allow the particle to travel
through different electric fields for the two systems in remote spacetime
regions (e.g. $E_{2}\left(  x,t^{\prime}\right)  \neq E_{1}\left(  x,t\right)
$ \ if $\ \ t^{\prime}<t$ \ etc.).

\bigskip

It is therefore clear that finding the appropriate $\ \Lambda(x,t)$ \ that
achieves the above mapping \textit{in full generality} will require a careful
solution of the system of PDEs (\ref{Basic11}), applied to only one spatial
variable, namely%

\begin{equation}
\frac{\partial\Lambda(x,t)}{\partial x}=A(x,t)\qquad and\qquad-\frac{1}%
{c}\frac{\partial\Lambda(x,t)}{\partial t}=\phi\left(  x,t\right)
\label{xt-BasicSystem}%
\end{equation}
(with $\ A(x,t)=A_{2}(x,t)-A_{1}(x,t)$ \ and $\ \phi\left(  x,t\right)
=\phi_{2}\left(  x,t\right)  -\phi_{1}\left(  x,t\right)  $). This system is
underdetermined in the sense that we only have knowledge of $\Lambda$ at an
initial point $(x_{0},t_{0})$ and with no further boundary conditions (hence
multiplicities of solutions being generally expected, see below). By following
a careful procedure of integrations\cite{jphysa} we finally obtain 2 distinct
solutions (depending on which equation we integrate first)$\boldsymbol{:}$ the
first solution is%

\begin{equation}
\Lambda(x,t)=\Lambda(x_{0},t_{0})+%
{\displaystyle\int\limits_{x_{0}}^{x}}
A(x^{\prime},t)dx^{\prime}-c%
{\displaystyle\int\limits_{t_{0}}^{t}}
\phi(x_{0},t^{\prime})dt^{\prime}+\left\{  c%
{\displaystyle\int\limits_{t_{0}}^{t}}
dt^{\prime}%
{\displaystyle\int\limits_{x_{0}}^{x}}
dx^{\prime}E(x^{\prime},t^{\prime})+g(x)\right\}  +\tau(t_{0})
\label{LambdaStatic1}%
\end{equation}
with $\ g(x)$ required to be chosen\ so that the quantity $\ \left\{  c%
{\displaystyle\int\limits_{t_{0}}^{t}}
dt^{\prime}%
{\displaystyle\int\limits_{x_{0}}^{x}}
dx^{\prime}E(x^{\prime},t^{\prime})+g(x)\right\}  $ \ is independent of $x$,
and (from an inverted route of integrations) the second solution turns out to be%

\begin{equation}
\Lambda(x,t)=\Lambda(x_{0},t_{0})+%
{\displaystyle\int\limits_{x_{0}}^{x}}
A(x^{\prime},t_{0})dx^{\prime}-c\int_{t_{0}}^{t}\phi\left(  x,t^{\prime
}\right)  dt^{\prime}+\left\{  -c%
{\displaystyle\int\limits_{x_{0}}^{x}}
dx^{\prime}%
{\displaystyle\int\limits_{t_{0}}^{t}}
dt^{\prime}E(x^{\prime},t^{\prime})+\hat{g}(t)\right\}  +\chi(x_{0})
\label{LambdaStatic2}%
\end{equation}
with $\ \hat{g}(t)$ \ to be chosen in such a way that \ $\left\{  -c%
{\displaystyle\int\limits_{x_{0}}^{x}}
dx^{\prime}%
{\displaystyle\int\limits_{t_{0}}^{t}}
dt^{\prime}E(x^{\prime},t^{\prime})+\hat{g}(t)\right\}  \ $\ is independent
of$\ t$.

\bigskip In the above $E=(E_{2}-E_{1})$ is the difference of electric fields
in the two systems, which can be nonvanishing at regions remote to the
observation point $(x,t)$ (see examples later below). (Note again that at the
point of observation $E(x,t)=0$, signifying the basic fact that the fields in
the two systems are identical at the observation point $(x,t)$). The constant
last terms in both solutions can be shown to be related to possible
multiplicities of $\Lambda$ (for a full discussion see \cite{jphysa}) and they
are zero in simple-connected spacetimes. Also note again that the integrations
of potentials in (\ref{LambdaStatic1}) and (\ref{LambdaStatic2}) indeed form
paths that continuously connect $(x_{0},t_{0})$ to $(x,t)$ in the $xt$-plane
(the red-arrow and green-arrow paths of Fig.1(a)), a property that the
incorrectly used solution (\ref{BrownHolland}) does \textit{not} have.

\bigskip By \textquotedblleft going backwards\textquotedblright\ one can
directly verify that (\ref{LambdaStatic1}) or (\ref{LambdaStatic2}) are indeed
solutions of the basic system of PDEs (\ref{xt-BasicSystem}), \textbf{even for
any nonzero} $E(x^{\prime},t^{\prime})$ \ (in regions $(x^{\prime},t^{\prime
})\neq(x,t)$). Indeed, if we call our first solution
\ (eq.(\ref{LambdaStatic1})) for simple-connected spacetime $\Lambda_{1}$, namely%

\begin{equation}
\Lambda_{1}(x,t)=\Lambda_{1}(x_{0},t_{0})+%
{\displaystyle\int\limits_{x_{0}}^{x}}
A(x^{\prime},t)dx^{\prime}-c%
{\displaystyle\int\limits_{t_{0}}^{t}}
\phi(x_{0},t^{\prime})dt^{\prime}+\left\{  c%
{\displaystyle\int\limits_{t_{0}}^{t}}
dt^{\prime}%
{\displaystyle\int\limits_{x_{0}}^{x}}
dx^{\prime}E(x^{\prime},t^{\prime})+g(x)\right\}  \label{L1}%
\end{equation}
with $g(x)$ chosen so that $\left\{  c%
{\displaystyle\int\limits_{t_{0}}^{t}}
dt^{\prime}%
{\displaystyle\int\limits_{x_{0}}^{x}}
dx^{\prime}E(x^{\prime},t^{\prime})+g(x)\right\}  $ is independent of $x$,
then we have (even for $E(x^{\prime},t^{\prime})\neq0$ for $(x^{\prime
},t^{\prime})\neq(x,t)$)$\boldsymbol{:}$

\bigskip

\textbf{A)} $\ \frac{\partial\Lambda_{1}(x,t)}{\partial x}=A(x,t)\qquad
$satisfied trivially\qquad$\checkmark$ \ \ \ \ \ \ \ \ \ \ \ 

(because the quantity $\left\{  ....\right\}  $ is independent of $x$).

\bigskip

\bigskip\textbf{B)} \ $-\frac{1}{c}\frac{\partial\Lambda_{1}(x,t)}{\partial
t}=-\frac{1}{c}%
{\displaystyle\int\limits_{x_{0}}^{x}}
\frac{\partial A(x^{\prime},t)}{\partial t}dx^{\prime}+\phi(x_{0},t)-%
{\displaystyle\int\limits_{x_{0}}^{x}}
E(x^{\prime},t)dx^{\prime}-\frac{1}{c}\frac{\partial g(x)}{\partial t}$
,\ \ \ \ \qquad

(the last term being trivially zero, $\frac{\partial g(x)}{\partial t}=0$ ),
and then with the substitution

$-\frac{1}{c}\frac{\partial A(x^{\prime},t)}{\partial t}=\frac{\partial
\phi(x^{\prime},t)}{\partial x^{\prime}}+E(x^{\prime},t)$

we obtain

$-\frac{1}{c}\frac{\partial\Lambda_{1}(x,t)}{\partial t}=%
{\displaystyle\int\limits_{x_{0}}^{x}}
\frac{\partial\phi(x^{\prime},t)}{\partial x^{\prime}}dx^{\prime}+%
{\displaystyle\int\limits_{x_{0}}^{x}}
E(x^{\prime},t)dx^{\prime}+\phi(x_{0},t)-%
{\displaystyle\int\limits_{x_{0}}^{x}}
E(x^{\prime},t)dx^{\prime}$. \ \ \ 

(i) We see that the 2nd and 4th terms of the rhs \textit{cancel each other}, and

(ii) the 1st term of the rhs is \ $%
{\displaystyle\int\limits_{x_{0}}^{x}}
\frac{\partial\phi(x^{\prime},t)}{\partial x^{\prime}}dx^{\prime}%
=\phi(x,t)-\phi(x_{0},t).\qquad$\ 

Hence finally

$-\frac{1}{c}\frac{\partial\Lambda_{1}(x,t)}{\partial t}=\phi(x,t).\qquad
\checkmark$

\bigskip We have directly shown therefore that the basic system of PDEs
(\ref{xt-BasicSystem}) is indeed satisfied by our \textbf{generalized}
solution $\Lambda_{1}(x,t),$ \textbf{even for any nonzero} $E(x^{\prime
},t^{\prime})$ \ (in regions $(x^{\prime},t^{\prime})\neq(x,t)$). (Once again
note, however, that at the point of observation $E(x,t)=0$, indicating the
essential fact that the fields in the two systems are equal (recall that
$E=E_{2}-E_{1}$) at the observation point $(x,t)$. It should be noted that the
function $g(x)$ owes its existence to the fact that the spacetime point of
observation $(x,t)$ is outside the $E$-distribution (hence the term
\textit{nonlocal,} used for the effect of the field-difference $E$ on the
phases), and the reader can clearly see this in the \textquotedblleft
striped\textquotedblright\ $E$-distributions of the examples that follow later
in this Section.

In a completely analogous way, one can easily see that our alternative
solution (eq.(\ref{LambdaStatic2})) also satisfies the basic system of PDEs
above. Indeed, if we call our second (alternative) solution
(eq.(\ref{LambdaStatic2})) for simple-connected spacetime $\Lambda_{2}$, namely%

\begin{equation}
\Lambda_{2}(x,t)=\Lambda_{2}(x_{0},t_{0})+%
{\displaystyle\int\limits_{x_{0}}^{x}}
A(x^{\prime},t_{0})dx^{\prime}-c\int_{t_{0}}^{t}\phi\left(  x,t^{\prime
}\right)  dt^{\prime}+\left\{  -c%
{\displaystyle\int\limits_{x_{0}}^{x}}
dx^{\prime}%
{\displaystyle\int\limits_{t_{0}}^{t}}
dt^{\prime}E(x^{\prime},t^{\prime})+\hat{g}(t)\right\}  \label{L2}%
\end{equation}
with $\hat{g}(t)$ chosen so that $\left\{  -c%
{\displaystyle\int\limits_{x_{0}}^{x}}
dx^{\prime}%
{\displaystyle\int\limits_{t_{0}}^{t}}
dt^{\prime}E(x^{\prime},t^{\prime})+\hat{g}(t)\right\}  $ is independent of
$t$, then we have (even for $E(x^{\prime},t^{\prime})\neq0$ for $(x^{\prime
},t^{\prime})\neq(x,t)$)$\boldsymbol{:}$

\bigskip

\textbf{A)} $\ -\frac{1}{c}\frac{\partial\Lambda_{2}(x,t)}{\partial t}%
=\phi(x,t)\qquad$satisfied trivially\qquad$\checkmark$ \ \ \ \ \ \ \ \ \ \ \ 

(because the quantity $\left\{  ....\right\}  $ is independent of $t$).

\bigskip

\bigskip\textbf{B)} \ $\frac{\partial\Lambda_{2}(x,t)}{\partial x}%
=A(x,t_{0})-c%
{\displaystyle\int\limits_{t_{0}}^{t}}
\frac{\partial\phi(x,t^{\prime})}{\partial x}dt^{\prime}-c%
{\displaystyle\int\limits_{t_{0}}^{t}}
E(x,t^{\prime})dt^{\prime}+\frac{\partial\hat{g}(t)}{\partial x}$
,\ \ \ \ \qquad

(the last term being trivially zero, $\frac{\partial\hat{g}(t)}{\partial x}=0$
), and then with the substitution

$\frac{\partial\phi(x,t^{\prime})}{\partial x}=-E(x,t^{\prime})-\frac{1}%
{c}\frac{\partial A(x,t^{\prime})}{\partial t^{\prime}}$

we obtain

$\frac{\partial\Lambda_{2}(x,t)}{\partial x}=A(x,t_{0})+c%
{\displaystyle\int\limits_{t_{0}}^{t}}
E(x,t^{\prime})dt^{\prime}+%
{\displaystyle\int\limits_{t_{0}}^{t}}
\frac{\partial A(x,t^{\prime})}{\partial t^{\prime}}dt^{\prime}-c%
{\displaystyle\int\limits_{t_{0}}^{t}}
E(x,t^{\prime})dt^{\prime}$. \ \ \ 

(i) We see that the 2nd and 4th terms of the rhs \textit{cancel each other}, and

(ii) the 3rd term of the rhs is \ $%
{\displaystyle\int\limits_{t_{0}}^{t}}
\frac{\partial A(x,t^{\prime})}{\partial t^{\prime}}dt^{\prime}%
=A(x,t)-A(x,t_{0}).\qquad$\ 

Hence finally

$\frac{\partial\Lambda_{2}(x,t)}{\partial x}=A(x,t).\qquad\checkmark$

\bigskip

Once again, all the above are true for any nonzero $E(x^{\prime},t^{\prime})$
(in regions $(x^{\prime},t^{\prime})\neq(x,t)$) for arbitrary analytical
dependence of the remote field-difference on its arguments.

\bigskip%

\begin{figure}[ptb]%
\centering
\includegraphics[
height=0.9703in,
width=3.2681in
]%
{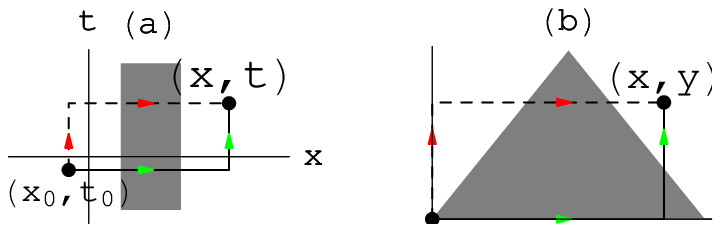}%
\end{figure}

\bigskip

\bigskip Let us now note that in (\ref{LambdaStatic1}) and
(\ref{LambdaStatic2}) the placement of $x_{0}$ and $t_{0}$ gives a
\textquotedblleft path-sense\textquotedblright\ to the line integrals in each
solution (each path consisting of 2 perpendicular line segments connecting
$(x_{0},t_{0})$ to $(x,t),$ with solution (\ref{LambdaStatic1}) having a
clockwise and solution (\ref{LambdaStatic2}) a counter-clockwise sense, see
red and green arrow paths in Fig.1)$\boldsymbol{;\ }$this way a natural
\textit{rectangle} is formed, within which the enclosed \textquotedblleft
electric fluxes\textquotedblright\ in spacetime appear to be crucial (showing
up as nonlocal contributions of the electric field-differences from regions
$(x^{\prime},t^{\prime})$ of space and time \textit{that are remote to the
observation point} $(x,t)$). These nonlocal terms in $\Lambda$ have a direct
effect on the wavefunction-phases at $(x,t)$. The actual manner in which this
happens is determined by the functions$\ g(x)$ or$\ \hat{g}(t)$--these must be
chosen in such a way as to satisfy their respective conditions in
(\ref{LambdaStatic1}) and (\ref{LambdaStatic2}). To see how these
functions$\ (g(x)$ or$\ \hat{g}(t))$ are actually determined, and what form
the above solutions take in nontrivial cases (and how they give new results,
i.e. \textit{not differing from the usual ones by a mere constant}) let us
first take examples of striped $E$-distributions in spacetime$\boldsymbol{:}$

\textbf{(a)} For the case of the extended \textit{vertical} strip (parallel to
the $t$-axis) of Fig.1(a) (the case of a one-dimensional capacitor that is
(arbitrarily and variably) charged for all time), then, for $x$ located
outside (and on the right of) the capacitor, the quantity$\ c%
{\displaystyle\int\limits_{t_{0}}^{t}}
dt^{\prime}%
{\displaystyle\int\limits_{x_{0}}^{x}}
dx^{\prime}E(x^{\prime},t^{\prime})$ in\ $\Lambda_{1}$ is \textit{already
independent of}$\ x$\ (since a displacement of the $(x,t)$-corner of the
rectangle to the right, along the $x$-direction, does not change the enclosed
\textquotedblleft electric flux\textquotedblright, see
Fig.1(a))$\boldsymbol{;}$ hence in this case the function $g(x)$ can be taken
as $g(x)=0$ (up to a constant $C$), because then the condition for $g(x)$
stated in the solution (\ref{LambdaStatic1}) (namely, that the quantity in
brackets must be independent of $x$) is indeed satisfied. (Note again that the
above $x$-independence of the enclosed \textquotedblleft electric
flux\textquotedblright\ is important for the existence of $g(x)$).

So for this setup, the nonlocal term in solution $\Lambda_{1}$
\textit{survives} (the quantity in brackets is nonvanishing), but \textit{it
is not constant}$\boldsymbol{:}$ this enclosed flux depends on $t$ (since the
enclosed flux \textit{does change} with a displacement of the $(x,t)$-corner
of the rectangle upwards, along the $t$-direction). Hence, by looking at the
alternative solution $\Lambda_{2}(x,t),$ the quantity$\ c%
{\displaystyle\int\limits_{x_{0}}^{x}}
dx^{\prime}%
{\displaystyle\int\limits_{t_{0}}^{t}}
dt^{\prime}E(x^{\prime},t^{\prime})$\ is dependent on$\ t$, so that $\hat
{g}(t)$ must be chosen as $\ \hat{g}(t)=+c%
{\displaystyle\int\limits_{x_{0}}^{x}}
dx^{\prime}%
{\displaystyle\int\limits_{t_{0}}^{t}}
dt^{\prime}E(x^{\prime},t^{\prime})$ (up to the same constant $C$)\ in order
to \textit{cancel} this $t$-dependence, so that its own condition stated in
the solution (\ref{LambdaStatic2}) (namely, that the quantity in brackets must
be independent of $t$) is indeed satisfied$\boldsymbol{;}$ as a result, the
quantity in brackets in solution $\Lambda_{2}$ disappears and there is no
nonlocal contribution in $\Lambda_{2}$ (for $C=0$). (If we had used a $C\neq
0$, the nonlocal contributions would be differently shared between the two
solutions, but without changing the Physics when we take the
\textit{difference} of the two solutions, see below).

With these choices of $\hat{g}(t)$ and $g(x)$, we already have new results
(compared to the standard ones of the integrals of potentials). I.e. one of
the two solutions, namely $\Lambda_{1}$ \textbf{is} affected nonlocally by the
enclosed flux (and this flux is \textbf{not} constant). Spelled out clearly,
the two results are:%

\begin{equation}
\Lambda_{1}(x,t)=\Lambda_{1}(x_{0},t_{0})+%
{\displaystyle\int\limits_{x_{0}}^{x}}
A(x^{\prime},t)dx^{\prime}-c%
{\displaystyle\int\limits_{t_{0}}^{t}}
\phi(x_{0},t^{\prime})dt^{\prime}+c%
{\displaystyle\int\limits_{t_{0}}^{t}}
dt^{\prime}%
{\displaystyle\int\limits_{x_{0}}^{x}}
dx^{\prime}E(x^{\prime},t^{\prime})+C \label{newsol1}%
\end{equation}

\begin{equation}
\Lambda_{2}(x,t)=\Lambda_{2}(x_{0},t_{0})+%
{\displaystyle\int\limits_{x_{0}}^{x}}
A(x^{\prime},t_{0})dx^{\prime}-c\int_{t_{0}}^{t}\phi\left(  x,t^{\prime
}\right)  dt^{\prime}+C. \label{newsol2}%
\end{equation}

And now it is easy to note that, if we subtract the two solutions $\Lambda
_{1}$ and $\Lambda_{2}$ (and of course assume, as usual, single-valuedness of
$\Lambda$ at the initial point $(x_{0},t_{0})$, i.e. $\Lambda_{1}(x_{0}%
,t_{0})=\Lambda_{2}(x_{0},t_{0})$) the result is \textit{zero} (i.e. it so
happens that the electric flux determined by the potential-integrals is
exactly cancelled by the nonlocal term of electric fields (i.e. the term that
survives in $\Lambda_{1}$ above)), a cancellation effect that is important and
that will be generalized in later Sections.

\textbf{(b) }In the \textquotedblleft dual case\textquotedblright\ of an
extended \textit{horizontal} strip - parallel to the $x$-axis (that
corresponds to a nonzero electric field in all space that has however a finite
duration $T)$, the proper choices (for observation time instant $t>T$) are
basically reverse (i.e. we can now take $\hat{g}(t)=0$ \ and $g(x)=-c%
{\displaystyle\int\limits_{t_{0}}^{t}}
dt^{\prime}%
{\displaystyle\int\limits_{x_{0}}^{x}}
dx^{\prime}E(x^{\prime},t^{\prime})$ (since the \textquotedblleft electric
flux\textquotedblright\ enclosed in the \textquotedblleft observation
rectangle\textquotedblright\ now depends on $x$, but not on $t$), with both
choices always up to a common constant) and once again we can easily see, upon
subtraction of the two solutions, a similar cancellation effect. In this case
again, the results are also new (a \ nonlocal term survives now in
$\Lambda_{2}$). Again spelled out clearly, these are:%

\begin{equation}
\Lambda_{1}(x,t)=\Lambda_{1}(x_{0},t_{0})+%
{\displaystyle\int\limits_{x_{0}}^{x}}
A(x^{\prime},t)dx^{\prime}-c%
{\displaystyle\int\limits_{t_{0}}^{t}}
\phi(x_{0},t^{\prime})dt^{\prime}+C \label{solsol1}%
\end{equation}

\begin{equation}
\Lambda_{2}(x,t)=\Lambda_{2}(x_{0},t_{0})+%
{\displaystyle\int\limits_{x_{0}}^{x}}
A(x^{\prime},t_{0})dx^{\prime}-c\int_{t_{0}}^{t}\phi\left(  x,t^{\prime
}\right)  dt^{\prime}-c%
{\displaystyle\int\limits_{x_{0}}^{x}}
dx^{\prime}%
{\displaystyle\int\limits_{t_{0}}^{t}}
dt^{\prime}E(x^{\prime},t^{\prime})+C \label{solsol2}%
\end{equation}
their difference also being zero.

\bigskip\textbf{(c) }And if we want cases that are more involved (with the
nonlocal contributions appearing nontrivially in \textbf{both} solutions
$\Lambda_{1}$ and $\Lambda_{2}$ and with $g(x)$ and $\hat{g}(t)$ not being
\textquotedblleft immediately visible\textquotedblright) we must again
consider different shapes of $E$-distribution. One such case (a triangular
$E$-distribution) is shown in Fig.1(b) (for the corresponding magnetic case to
be discussed in the next Section, which however is completely analogous); in
this case the enclosed flux depends on \textit{both} $x$ and $t$ (but can be
shown to be separable, so that the functions $g(x)$ and $\hat{g}(t)$ still
exist and can easily be found in closed analytical form, see next Section). As
for the last constant terms $\tau(t_{0})$ and $\chi(x_{0})$ (what we will call
\textquotedblleft multiplicities\textquotedblright), these are only present
(nonvanishing) when $\Lambda$ is expected to be multivalued, i.e. in cases of
motion in multiple-connected spacetimes, and are then related to the fluxes in
the inaccessible regions: in the electric Aharonov-Bohm setup, the prototype
of multiple-connectivity in spacetime\cite{Iddings}, it turns out\cite{jphysa}
that $\tau(t_{0})=-\chi(x_{0})=$ enclosed (and here inaccessible)
\textquotedblleft electric flux\textquotedblright, and if these values are
substituted in (\ref{LambdaStatic1}) and (\ref{LambdaStatic2}) they cancel out
the new nonlocal terms and lead to the usual electric Aharonov-Bohm result (of
mere integrals over potentials). As for other, more esoteric properties of the
new solutions in simple-connected spacetimes, it can be rigorously
shown\cite{jphysa} that solutions (\ref{LambdaStatic1}) and
(\ref{LambdaStatic2}) are actually \textit{equal}, because $g(x)$ turns out to
be equal to the $t$-independent bracket of (\ref{LambdaStatic2}), and $\hat
{g}(t)$ turns out to be equal to the $x$-independent bracket of
(\ref{LambdaStatic1}), the nonlocal terms having therefore the tendency to
exactly cancel the \textquotedblleft Aharonov-Bohm terms\textquotedblright%
\ (this being true \textit{for arbitrary shapes and analytical form of
}$E(x,t)$).

\section{2-D Static Case}

\bigskip After having discussed fully the simple $(x,t)$-case, let us for
completeness give the analogous (Euclidian-rotated in 4-D spacetime)
derivation for $(x,y)$-variables and briefly discuss the properties of the
simpler static solutions in spatial two-dimensionality. We will simply need to
apply the same methodology (of solution of a system of PDEs) to such static
spatially two-dimensional cases (so that now different (remote) magnetic
fields for the two systems, perpendicular to the 2-D space, will arise). For
such cases we need to solve the system of PDEs that is now of the form%

\begin{equation}
\frac{\partial\Lambda(x,y)}{\partial x}=A_{x}(x,y)\qquad and\qquad
\frac{\partial\Lambda(x,y)}{\partial y}=A_{y}(x,y). \label{usualgrad}%
\end{equation}
By following then a similar procedure of integrations\cite{jphysa} we obtain
the following general solution%
\begin{equation}
\Lambda(x,y)=\Lambda(x_{0},y_{0})+%
{\displaystyle\int\limits_{x_{0}}^{x}}
A_{x}(x^{\prime},y)dx^{\prime}+%
{\displaystyle\int\limits_{y_{0}}^{y}}
A_{y}(x_{0},y^{\prime})dy^{\prime}+\left\{
{\displaystyle\int\limits_{y_{0}}^{y}}
dy^{\prime}%
{\displaystyle\int\limits_{x_{0}}^{x}}
dx^{\prime}B_{z}(x^{\prime},y^{\prime})+g(x)\right\}  +f(y_{0})
\label{Lambda(x,y)new}%
\end{equation}

\[
with\text{ \ }g(x)\text{ \ }chosen\ \ so\text{ \ }that\text{ \ }\left\{
{\displaystyle\int\limits_{y_{0}}^{y}}
dy^{\prime}%
{\displaystyle\int\limits_{x_{0}}^{x}}
dx^{\prime}B_{z}(x^{\prime},y^{\prime})+g(x)\right\}  \boldsymbol{:}\text{ is
}\mathsf{independent\ of\ }\ x,
\]
a result that applies to cases where the particle goes through
\textit{different} perpendicular magnetic fields $\boldsymbol{B}_{2}$ and
$\boldsymbol{B}_{1}$ \textit{in spatial regions that are remote to (i.e. do
not contain) the observation point} $(x,y)$ (and in the above $B_{z}%
={\huge (}\boldsymbol{B}_{2}-\boldsymbol{B}_{1}{\huge )}_{z}$). The reader
should note that the first 3 terms of (\ref{Lambda(x,y)new}) are the (total)
Dirac phase along two perpendicular segments that continuously connect the
initial point $(x_{0},y_{0})$ to the point of observation $(x,y)$, \textit{in
a clockwise sense }(see for example the red-arrow paths in Fig.1(b)). But
apart from this Dirac phase, we also have nonlocal contributions from $B_{z}$
and its flux within the \textquotedblleft observation
rectangle\textquotedblright\ (see i.e. the rectangle being formed by the red-
and green-arrow paths in Fig.1(b)). Below we will directly verify that
(\ref{Lambda(x,y)new}) is indeed a solution of (\ref{usualgrad}) (even for
$B_{z}(x^{\prime},y^{\prime})\neq0$ for $(x^{\prime},y^{\prime})\neq(x,y)$).
Alternatively, by following the reverse route of integrations, we finally
obtain the following alternative general solution%

\begin{equation}
\Lambda(x,y)=\Lambda(x_{0},y_{0})+\int_{x_{0}}^{x}A_{x}(x^{\prime}%
,y_{0})dx^{\prime}+\int_{y_{0}}^{y}A_{y}(x,y^{\prime})dy^{\prime}+\left\{
{\Huge -}%
{\displaystyle\int\limits_{x_{0}}^{x}}
dx^{\prime}%
{\displaystyle\int\limits_{y_{0}}^{y}}
dy^{\prime}B_{z}(x^{\prime},y^{\prime})+h(y)\right\}  +\hat{h}(x_{0})
\label{Lambda(x,y)4}%
\end{equation}

\[
with\text{ \ }h(y)\text{ \ }chosen\text{ \ }so\text{ \ }that\text{ \ }\left\{
{\Huge -}%
{\displaystyle\int\limits_{x_{0}}^{x}}
dx^{\prime}%
{\displaystyle\int\limits_{y_{0}}^{y}}
dy^{\prime}B_{z}(x^{\prime},y^{\prime})+h(y)\right\}  \boldsymbol{:}\text{ is
}\mathsf{independent\ of\ }\ y,
\]
and again the reader should note that, apart from the first 3 terms (the
(total) Dirac phase along the two other (alternative) perpendicular segments
(connecting $(x_{0},y_{0})$ to $(x,y)$), now \textit{in a counterclockwise
sense }(the green-arrow paths in Fig.1(b))), we also have nonlocal
contributions from the flux of $B_{z}$ that is enclosed within the same
\textquotedblleft observation rectangle\textquotedblright\ (that is naturally
defined by the four segments of the two solutions (Fig.1(b))).

In all the above, $A_{x}$ and $A_{y}$ are the Cartesian components of
$\ \boldsymbol{A}\mathbf{(r)}=\boldsymbol{A}(x,y)=\boldsymbol{A}%
_{\boldsymbol{2}}(\mathbf{r})-\boldsymbol{A}_{\boldsymbol{1}}(\mathbf{r})$,
and, as already mentioned, \ $B_{z}$ \ is the difference between
(perpendicular) magnetic fields that the two systems may experience in regions
that \textit{do not contain} the observation point $(x,y)$ (i.e.
$B_{z}(x^{\prime},y^{\prime})={\huge (}\boldsymbol{B}_{\boldsymbol{2}%
}(x^{\prime},y^{\prime})-\boldsymbol{B}_{\boldsymbol{1}}(x^{\prime},y^{\prime
}){\huge )}_{z}=\frac{\partial A_{y}(x^{\prime},y^{\prime})}{\partial
x^{\prime}}-$ $\frac{\partial A_{x}(x^{\prime},y^{\prime})}{\partial
y^{\prime}}$, \ and, although at the point of observation $(x,y)$ we have
$B_{z}(x,y)=0$ (already emphasized in the Introductory Sections), this
$B_{z}(x^{\prime},y^{\prime})$ can be nonzero for \ $(x^{\prime},y^{\prime
})\neq(x,y)$). It should be noted that it is because of $B_{z}(x,y)=0$ that
the functions $g(x)$ and $h(y)$ of (\ref{Lambda(x,y)new}) and
(\ref{Lambda(x,y)4}) can be found, and the new solutions therefore exist (and
are nontrivial). For the impatient reader, simple physical examples with the
associated analytical forms of $g(x)$ and $h(y)$ (derived in detail) are given
later in this Section.

One can again show that the 2 solutions are equal for simple-connected space
(when the last constant terms $f(y_{0})$ and $\hat{h}(x_{0})$ are vanishing),
and for multiple-connectivity the values of the multiplicities $f(y_{0})$ and
$\hat{h}(x_{0})$ cancel out the nonlocalities and reduce the above to the
usual result of mere $A$-integrals along the 2 paths (i.e. two simple Dirac phases).

A direct \textquotedblleft backwards\textquotedblright\ verification that
(\ref{Lambda(x,y)new}) and (\ref{Lambda(x,y)4}) do indeed satisfy the basic
system (\ref{usualgrad}) \textbf{(even for cases with }$\boldsymbol{B}%
_{z}\boldsymbol{\neq0}$ in remote regions of space\textbf{) }can be made along
similar lines to the ones of the last Section$\boldsymbol{:}$ for
simple-connected space, let us call our solution (\ref{Lambda(x,y)new})
$\Lambda_{3}$, namely%

\begin{equation}
\Lambda_{3}(x,y)=\Lambda_{3}(x_{0},y_{0})+%
{\displaystyle\int\limits_{x_{0}}^{x}}
A_{x}(x^{\prime},y)dx^{\prime}+%
{\displaystyle\int\limits_{y_{0}}^{y}}
A_{y}(x_{0},y^{\prime})dy^{\prime}+\left\{
{\displaystyle\int\limits_{y_{0}}^{y}}
dy^{\prime}%
{\displaystyle\int\limits_{x_{0}}^{x}}
dx^{\prime}B_{z}(x^{\prime},y^{\prime})+g(x)\right\}  \label{L3}%
\end{equation}
with\ $g(x)$ chosen so that\ $\ \left\{
{\displaystyle\int\limits^{y}}
{\displaystyle\int\limits^{x}}
B_{z}+g(x)\right\}  \boldsymbol{\ }$is independent of $x.$ We then have (even
for $B_{z}(x^{\prime},y^{\prime})\neq0$ for $(x^{\prime},y^{\prime})\neq
(x,y)$)$\boldsymbol{:}$

\textbf{A) \ }$\frac{\partial\Lambda_{3}(x,y)}{\partial x}=A_{x}(x,y)\qquad
$satisfied trivially\qquad$\checkmark$

(because the quantity $\left\{  ...\right\}  $ is independent of $x$).

\textbf{B) \ }$\frac{\partial\Lambda_{3}(x,y)}{\partial y}=%
{\displaystyle\int\limits_{x_{0}}^{x}}
\frac{\partial A_{x}(x^{\prime},y)}{\partial y}dx^{\prime}+A_{y}(x_{0},y)+%
{\displaystyle\int\limits_{x_{0}}^{x}}
B_{z}(x^{\prime},y)dx^{\prime}+\frac{\partial g(x)}{\partial y},$

(the last term being trivially zero, $\frac{\partial g(x)}{\partial y}=0$),
and then with the substitution

$\frac{\partial A_{x}(x^{\prime},y)}{\partial y}=\frac{\partial A_{y}%
(x^{\prime},y)}{\partial x^{\prime}}-B_{z}(x^{\prime},y)$

we obtain

$\frac{\partial\Lambda_{3}(x,y)}{\partial y}=%
{\displaystyle\int\limits_{x_{0}}^{x}}
\frac{\partial A_{y}(x^{\prime},y)}{\partial x^{\prime}}dx^{\prime}-%
{\displaystyle\int\limits_{x_{0}}^{x}}
B_{z}(x^{\prime},y)dx^{\prime}+A_{y}(x_{0},y)+%
{\displaystyle\int\limits_{x_{0}}^{x}}
B_{z}(x^{\prime},y)dx^{\prime}.$

(i) We see that the 2nd and 4th terms of the right-hand-side (rhs)
\textit{cancel each other}, and

(ii) the 1st term of the rhs is $%
{\displaystyle\int\limits_{x_{0}}^{x}}
\frac{\partial A_{y}(x^{\prime},y)}{\partial x^{\prime}}dx^{\prime}%
=A_{y}(x,y)-A_{y}(x_{0},y).$

Hence finally

$\frac{\partial\Lambda_{3}(x,y)}{\partial y}=A_{y}(x,y).\qquad\checkmark$

\bigskip

We have directly shown therefore (by \textquotedblleft going
backwards\textquotedblright) that the basic system of PDEs (\ref{usualgrad})
is indeed satisfied by our generalized solution $\Lambda_{3}(x,y),$ even for
any nonzero $B_{z}(x^{\prime},y^{\prime})$\ (in regions $(x^{\prime}%
,y^{\prime})\neq(x,y)\boldsymbol{;}$ recall that always $B_{z}(x,y)=0$). To
fully appreciate the above simple proof, the reader is again urged to look at
the cases of \textquotedblleft striped\textquotedblright\ $B_{z}%
$-distributions later below, the point of observation $(x,y)$ always lying
outside the strips, so that the above function $g(x)$ can easily be
determined, and the new solutions really \textit{exist }- and they are nontrivial.

In a completely analogous way, one can easily see that our alternative
solution (eq.(\ref{Lambda(x,y)4})) also satisfies the basic system of PDEs
above. Indeed, if we call this second static solution $\Lambda_{4}$, namely%

\begin{equation}
\Lambda_{4}(x,y)=\Lambda_{4}(x_{0},y_{0})+\int_{x_{0}}^{x}A_{x}(x^{\prime
},y_{0})dx^{\prime}+\int_{y_{0}}^{y}A_{y}(x,y^{\prime})dy^{\prime}+\left\{
{\Huge -}%
{\displaystyle\int\limits_{x_{0}}^{x}}
dx^{\prime}%
{\displaystyle\int\limits_{y_{0}}^{y}}
dy^{\prime}B_{z}(x^{\prime},y^{\prime})+h(y)\right\}  \label{L4}%
\end{equation}
with $h(y)$ chosen so that\ $\ \left\{  {\Huge -}%
{\displaystyle\int\limits^{x}}
{\displaystyle\int\limits^{y}}
B_{z}+h(y)\right\}  \boldsymbol{:}$ is independent of\textsf{ }$y$, then we
have (even for $B_{z}(x^{\prime},y^{\prime})\neq0$ for $(x^{\prime},y^{\prime
})\neq(x,y)$)$\boldsymbol{:}$

\bigskip

\textbf{A) \ }$\frac{\partial\Lambda_{4}(x,y)}{\partial y}=A_{y}(x,y)\qquad
$satisfied trivially\qquad$\checkmark$

(because the quantity $\left\{  ...\right\}  $ is independent of $y$).

\bigskip

\textbf{B) \ }$\frac{\partial\Lambda_{4}(x,y)}{\partial x}=A_{x}(x,y_{0})+%
{\displaystyle\int\limits_{y_{0}}^{y}}
\frac{\partial A_{y}(x,y^{\prime})}{\partial x}dy^{\prime}-%
{\displaystyle\int\limits_{y_{0}}^{y}}
B_{z}(x,y^{\prime})dy^{\prime}+\frac{\partial h(y)}{\partial x},$

(the last term being trivially zero, $\frac{\partial h(y)}{\partial x}=0$),
and then with the substitution

$\frac{\partial A_{y}(x,y^{\prime})}{\partial x}=\frac{\partial A_{x}%
(x,y^{\prime})}{\partial y^{\prime}}+B_{z}(x,y^{\prime})$

we obtain

$\frac{\partial\Lambda_{4}(x,y)}{\partial x}=A_{x}(x,y_{0})+%
{\displaystyle\int\limits_{y_{0}}^{y}}
\frac{\partial A_{x}(x,y^{\prime})}{\partial y^{\prime}}dy^{\prime}+%
{\displaystyle\int\limits_{y_{0}}^{y}}
B_{z}(x,y^{\prime})dy^{\prime}-%
{\displaystyle\int\limits_{y_{0}}^{y}}
B_{z}(x,y^{\prime})dy^{\prime}.$

(i) We see that the last two terms of the rhs \textit{cancel each other}, and

(ii) the 2nd term of the rhs is $%
{\displaystyle\int\limits_{y_{0}}^{y}}
\frac{\partial A_{x}(x,y^{\prime})}{\partial y^{\prime}}dy^{\prime}%
=A_{x}(x,y)-A_{x}(x,y_{0}).$

Hence finally

$\frac{\partial\Lambda_{4}(x,y)}{\partial x}=A_{x}(x,y).\qquad\checkmark$

\bigskip Once again, all the above are true for any nonzero $B_{z}(x^{\prime
},y^{\prime})$ (in regions $(x^{\prime},y^{\prime})\neq(x,y)$) for arbitrary
analytical dependence of the remote field-difference on its arguments. And for
a clearer understanding of this proof let us now turn to the \textquotedblleft
striped\textquotedblright\ examples promised earlier.

In order to see again how the above solutions appear in nontrivial cases (and
how they give completely new results, i.e. \textit{not differing from the
usual ones }(\textit{i.e. from the Dirac phase})\textit{ by a mere constant})
let us first take examples of striped $B_{z}$-distributions in
space$\boldsymbol{:}$

\bigskip

\bigskip

\textbf{(a)} For the case of an extended \textit{vertical} strip - parallel to
the $y$-axis, such as in Fig.1(a) (imagine $t$ replaced by $y$) (i.e. for the
case that the particle has actually passed through nonzero $B_{z}$, hence
through \textit{different} magnetic fields in the two (mapped) systems), then,
for $x$ located outside (and on the right of) the strip, the quantity$\
{\displaystyle\int\limits_{y_{0}}^{y}}
dy^{\prime}%
{\displaystyle\int\limits_{x_{0}}^{x}}
dx^{\prime}B_{z}(x^{\prime},y^{\prime})$ in\ $\Lambda_{3}$ \textit{is already
independent of}$\ x$\ (since a displacement of the $(x,y)$-corner of the
rectangle to the right, along the $x$-direction, does not change the enclosed
magnetic flux $-$ see Fig. 1(a) for the analogous $(x,t)$-case discussed
earlier)$\boldsymbol{.}$ Indeed, in this case the above quantity (the enclosed
flux within the \textquotedblleft observation rectangle\textquotedblright)
does not depend on the $x$-position of the observation point, but on the
positioning of the boundaries of the $B_{z}$-distribution in the $x$-direction
(better, on the constant width of the strip) $-$ as the $x$-integral does not
give any further contribution when the dummy variable $x^{\prime}$ goes out of
the strip. In fact, in this case the enclosed flux depends on $y$ as we
discuss below (but, again, not on $x$). Hence, for this case, the function
$g(x)$ can be easily determined: it can be taken as $g(x)=0$ (up to a constant
$C$), because then the condition for $g(x)$ stated in solution
(\ref{Lambda(x,y)new}) (namely, that the quantity in brackets must be
independent of $x$) is indeed satisfied.

We see therefore above that for this setup, the nonlocal term in solution
$\Lambda_{3}$ \textit{survives} (the quantity in brackets is nonvanishing),
but \textit{it is not constant}$\boldsymbol{:}$ as already noted, this
enclosed flux depends on $y$ (since the enclosed flux \textit{does change}
with a displacement of the $(x,y)$-corner of the rectangle upwards, along the
$y$-direction, as the $y$-integral \textit{is} affected by the positioning of
$y$ $-$ the higher the positioning of the observation point the more flux is
enclosed inside the observation rectangle). Hence, by looking at the
alternative solution $\Lambda_{4}(x,y),$ the quantity$\
{\displaystyle\int\limits_{x_{0}}^{x}}
dx^{\prime}%
{\displaystyle\int\limits_{y_{0}}^{y}}
dy^{\prime}B_{z}(x^{\prime},y^{\prime})$\ is \textit{dependent on}$\ y$, so
that $h(y)$ must be chosen as $\ h(y)=+%
{\displaystyle\int\limits_{x_{0}}^{x}}
dx^{\prime}%
{\displaystyle\int\limits_{y_{0}}^{y}}
dy^{\prime}B_{z}(x^{\prime},y^{\prime})$ (up to the same constant $C$)\ in
order to \textit{cancel} this $y$-dependence, so that its own condition stated
in solution (\ref{Lambda(x,y)4}) (namely, that the quantity in brackets must
be independent of $y$) is indeed satisfied$\boldsymbol{;}$ as a result, the
quantity in brackets in solution $\Lambda_{4}$ disappears and there is no
nonlocal contribution in $\Lambda_{4}$ (for $C=0$). (If we had used a $C\neq
0$, the nonlocal contributions would be shared between the two solutions in a
different manner, but without changing the Physics when we take the
\textit{difference} of the two solutions (see below)). [The crucial point in
the above for the existence of $g(x)$ and $h(y)$ is, once again, the fact that
$B_{z}=0$ at $(x,y)$, combined with the sharp boundaries of the nonvanishing
$B_{z}$-region].

With these choices of $h(y)$ and $g(x)$, we already have new results (compared
to the standard ones of the integrals of potentials). I.e. one of the two
solutions, namely $\Lambda_{3}$ \textbf{is} affected nonlocally by the
enclosed flux (and this flux is \textbf{not} constant). Spelled out clearly,
the two results are:%

\begin{equation}
\Lambda_{3}(x,y)=\Lambda_{3}(x_{0},y_{0})+%
{\displaystyle\int\limits_{x_{0}}^{x}}
A_{x}(x^{\prime},y)dx^{\prime}+%
{\displaystyle\int\limits_{y_{0}}^{y}}
A_{y}(x_{0},y^{\prime})dy^{\prime}+%
{\displaystyle\int\limits_{y_{0}}^{y}}
dy^{\prime}%
{\displaystyle\int\limits_{x_{0}}^{x}}
dx^{\prime}B_{z}(x^{\prime},y^{\prime})+C \label{sol3}%
\end{equation}

\begin{equation}
\Lambda_{4}(x,y)=\Lambda_{4}(x_{0},y_{0})+\int_{x_{0}}^{x}A_{x}(x^{\prime
},y_{0})dx^{\prime}+\int_{y_{0}}^{y}A_{y}(x,y^{\prime})dy^{\prime}+C.
\label{sol4}%
\end{equation}
And now it is easy to note that, if we subtract the two solutions $\Lambda
_{3}$ and $\Lambda_{4}$, the result is \textit{zero} (because the line
integrals of the vector potential $\boldsymbol{A}$ in the two solutions are in
opposite senses in the $(x,y)$ plane, hence their difference leads to a
\textit{closed} line integral of $\boldsymbol{A}$, which is in turn equal to
the enclosed magnetic flux, and this flux always happens to be of opposite
sign from that of the enclosed flux that explicitly appears as a
nonlocal\ contribution of the $B_{z}$-fields (i.e. the term that survives in
$\Lambda_{3}$ above)$\boldsymbol{.}$ Hence, the two solutions are
\textit{equal}. [We of course everywhere assumed, as usual, single-valuedness
of $\Lambda$ at the initial point $(x_{0},y_{0})$, i.e. $\Lambda_{1}%
(x_{0},y_{0})=\Lambda_{2}(x_{0},y_{0});$ matters of multivaluedness of
$\Lambda$ at the observation point $(x,y)$ will be addressed later below].

It is interesting that, formally speaking, the above \textit{equality} of the
two solutions is due to the fact that the $x$-independent quantity in brackets
of the 1st solution (\ref{Lambda(x,y)new}) is equal to the function $h(y)$ of
the 2nd solution (\ref{Lambda(x,y)4}), and the $y$-independent quantity in
brackets of the 2nd solution (\ref{Lambda(x,y)4}) is equal to the function
$g(x)$ of the first solution (\ref{Lambda(x,y)new}). This turns out to be a
general behavioral pattern of the two solutions in simple-connected space,
that is valid for any shape (and any analytical form) of $B_{z}$-distribution.
But most importantly, it should be noted that this vanishing of $\Lambda
_{3}(x,y)-\Lambda_{4}(x,y)$ is a cancellation effect that is emphasized
further later below and discussed in completely \textit{physical terms}.

\textbf{(b) }In the \textquotedblleft dual case\textquotedblright\ of an
extended \textit{horizontal} strip - parallel to the $x$-axis, the proper
choices (for $y$ above the strip) are basically reverse (i.e. we can now take
$h(y)=0$ \ and $g(x)=-%
{\displaystyle\int\limits_{y_{0}}^{y}}
dy^{\prime}%
{\displaystyle\int\limits_{x_{0}}^{x}}
dx^{\prime}B_{z}(x^{\prime},y^{\prime})$ $\ $(since the flux enclosed in the
rectangle now depends on $x$, but not on $y$), with both choices always up to
a common constant) and once again we can easily see, upon subtraction of the
two solutions, a similar cancellation effect. In this case as well, the
results are again new (a \ nonlocal term survives now in $\Lambda_{4}$). Again
spelled out clearly, these are:%

\begin{equation}
\Lambda_{3}(x,y)=\Lambda_{3}(x_{0},y_{0})+%
{\displaystyle\int\limits_{x_{0}}^{x}}
A_{x}(x^{\prime},y)dx^{\prime}+%
{\displaystyle\int\limits_{y_{0}}^{y}}
A_{y}(x_{0},y^{\prime})dy^{\prime}+C \label{solsol3}%
\end{equation}

\begin{equation}
\Lambda_{4}(x,y)=\Lambda_{4}(x_{0},y_{0})+\int_{x_{0}}^{x}A_{x}(x^{\prime
},y_{0})dx^{\prime}+\int_{y_{0}}^{y}A_{y}(x,y^{\prime})dy^{\prime}{\Huge -}%
{\displaystyle\int\limits_{x_{0}}^{x}}
dx^{\prime}%
{\displaystyle\int\limits_{y_{0}}^{y}}
dy^{\prime}B_{z}(x^{\prime},y^{\prime})+C \label{solsol4}%
\end{equation}
(their difference also being zero). Again here the crucial point is that,
because the $B_{z}$-configuration does \textit{not contain the point} $(x,y)$,
a displacement of this observation point upwards does \textit{not} change the
flux inside the \textquotedblleft observation rectangle\textquotedblright%
$\boldsymbol{;}$ this makes the new solutions (i.e the functions $g(x)$ and
$h(y)$) exist.

\bigskip

\textbf{(c) }And if we want cases that are more involved (i.e. with the
nonlocal contributions appearing nontrivially in \textbf{both} solutions
$\Lambda_{3}$ and $\Lambda_{4}$ and with $g(x)$ and $h(y)$ not being
\textquotedblleft immediately visible\textquotedblright), we must consider
different shapes of $B_{z}$-distributions. One such case is a triangular one
that is shown in Fig.1(b) (for simplicity an equilateral triangle, and with
the initial point $(x_{0},y_{0})=(0,0)$) and with the point of observation
$(x,y)$ being fairly close to the triangle's right side as in the figure. Note
that for such a configuration, the part of the magnetic flux that is inside
the \textquotedblleft observation rectangle\textquotedblright\ (defined by the
right upper corner $(x,y)$) depends on \textbf{both} $x$ \textbf{and} $y$. It
turns out, however, that this $(x$ and $y)-$dependent enclosed flux can be
written as a sum of separate $x$- and $y$-contributions, so that appropriate
$g(x)$ and $h(y)$ can still be found (each one of them must be chosen so that
it only cancels the corresponding variable's dependence of the enclosed flux).
For a homogeneous $B_{z}$ it is a rather straightforward exercise to determine
this enclosed part, i.e. the common area between the observation rectangle and
the equilateral triangle, and from this we can find the appropriate $g(x)$
that will cancel the $x$-dependence, and the appropriate $h(y)$ that will
cancel the $y$-dependence. These appropriate choices turn out to be%

\begin{equation}
g(x)=B_{z}\left[  \mathbf{-(}\sqrt{3}ax-\frac{\sqrt{3}}{2}x^{2})+\frac
{\sqrt{3}}{4}a^{2}\right]  +C \label{triangular1}%
\end{equation}
and%

\begin{equation}
h(y)=B_{z}\left[  \mathbf{(}ay-\frac{y^{2}}{\sqrt{3}})-\frac{\sqrt{3}}{4}%
a^{2}\right]  +C \label{triangular2}%
\end{equation}
with $a$ being the side of the equilateral triangle. We should emphasize that
expressions (\ref{triangular1}) and (\ref{triangular2}), if combined with
(\ref{Lambda(x,y)new}) or (\ref{Lambda(x,y)4}), give the nontrivial nonlocal
contributions of the field-difference $B_{z}$ of the remote magnetic fields on
$\Lambda$ of each solution (hence on the phase of the wavefunction of each
wavepacket travelling along each path) at the observation point $(x,y)$, that
always lies outside the $B_{z}$-triangle. (We mention again that in the case
of completely spread-out states, the equality of the two solutions at the
observation point essentially demonstrates the uniqueness (single-valuedness)
of the phase in simple-connected space). Further physical discussion of the
above cancellations, and a semiclassical interpretation, is given later below.

Finally, in more \textquotedblleft difficult\textquotedblright\ geometries,
i.e. when the shape of the $B_{z}$-distribution is such that the enclosed flux
does \textit{not}\textbf{ }decouple in a sum of separate $x$- and
$y$-contributions, \ such as cases of circularly shaped $B_{z}$-distributions,
it is advantageous to solve the system (\ref{usualgrad}) directly in
non-Cartesian (i.e. polar) coordinates\cite{jphysa}. A general comment that
can be made for general shapes is that, depending on the geometry of shape of
the $B_{z}$-distribution, an appropriate change of variables (to a new
coordinate system) may first be needed, so that generalized solutions of the
system (\ref{usualgrad}) can be found (namely, so that the enclosed flux
inside the \textit{transformed} observation rectangle (i.e. a slice of an
annular section in the case of polar coordinates) can be written as a sum of
separate (transformed) variables), and then the same methodology (as in the
above Cartesian cases) can be followed.

As we saw in the above examples, in case of a striped-distribution of the
magnetic field difference $B_{z}$, the functions $g(x)$ and $h(y)$ in
(\ref{Lambda(x,y)new}) and (\ref{Lambda(x,y)4}) have to be chosen in ways that
are compatible with their corresponding constraints (stated after
(\ref{Lambda(x,y)new}) and (\ref{Lambda(x,y)4})) and are completely analogous
to the above discussed $(x,t)$-cases$\boldsymbol{.}$ By then taking the
\textit{difference} of (\ref{Lambda(x,y)new}) and (\ref{Lambda(x,y)4})) we
obtain that the \textquotedblleft Aharonov-Bohm phase\textquotedblright\ (the
one originating from the \textit{closed} line integral of $A$'s) is exactly
cancelled by the additional nonlocal term of the magnetic fields (that the
particle passed through). This is quite reminiscent of the cancellation of
phases observed in the early experiments of Werner \& Brill\cite{WernerBrill}
for particles passing through a magnetic field (a cancellation between the
\textquotedblleft Aharonov-Bohm phase\textquotedblright\ and the semiclassical
phase picked up by the trajectories), and our method seems to provide a
natural explanation$\boldsymbol{:}$ as our results are general (and for
delocalized states in simple-connected space they basically demonstrate the
uniqueness of $\Lambda$), they are also valid and applicable to states that
describe wavepackets in classical motion, as \textit{was} the case of the
Werner \& Brill experiments. (A similar cancellation of an electric
Aharonov-Bohm phase, that has never been noted in the literature, also occurs
for particles passing through a static electric field, and this we will
independently prove below, again for semiclassical states). We conclude that,
for static cases, and when particles pass through nonvanishing classical
fields, the new nonlocal terms reported in the present work lead quite
generally to a cancellation of Aharonov-Bohm phases that had earlier been only
sketchily noticed and only at the semiclassical (magnetic) level.%

\begin{figure}[ptb]%
\centering
\includegraphics[
height=4.7651in,
width=6.365in
]%
{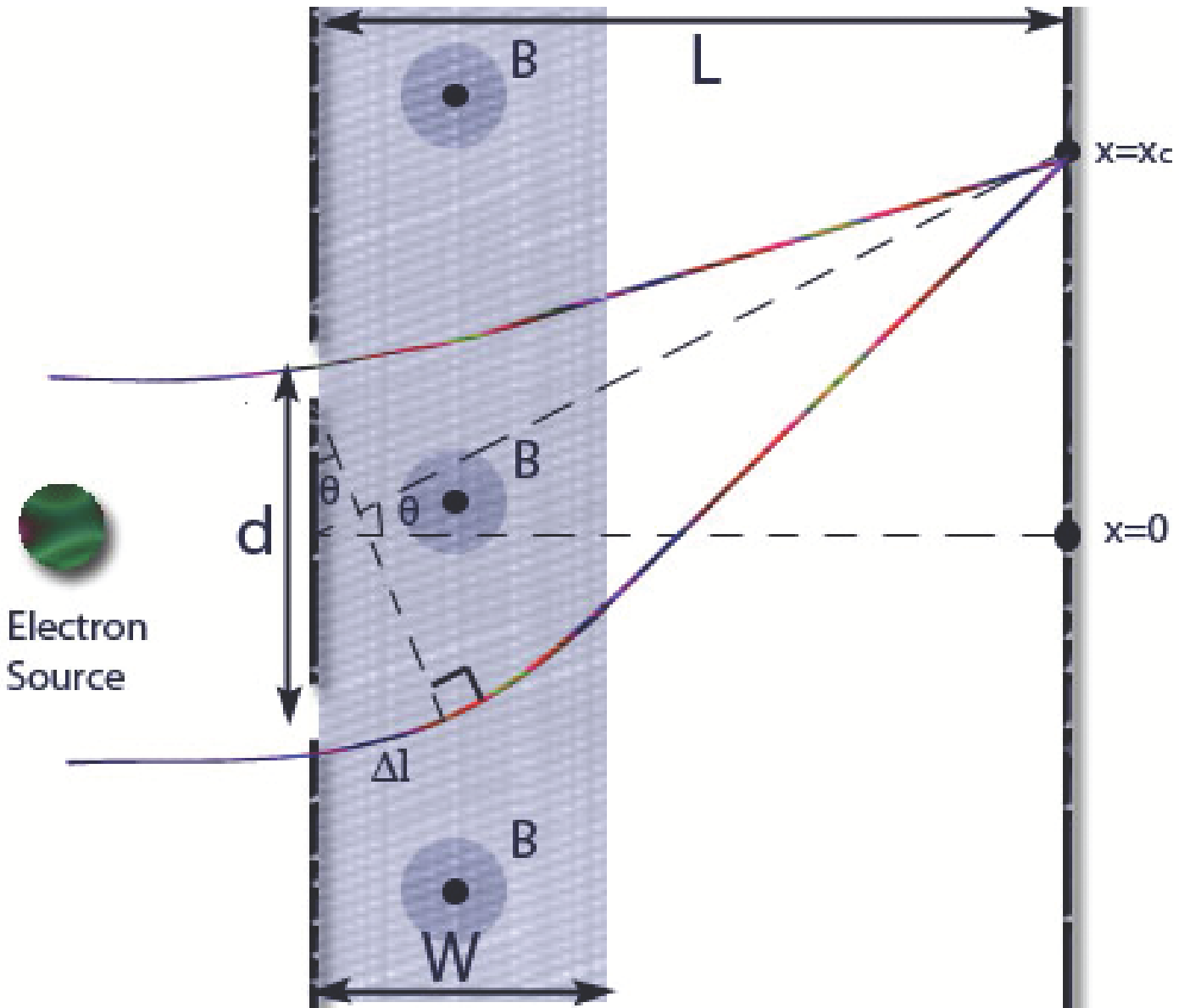}%
\end{figure}

\bigskip

What is however more important to point out here is that the above
cancellations can be understood as a compatibility between the Aharonov-Bohm
fringe-displacement and the trajectory-deflection due to the Lorentz force
(i.e. the semiclassical phase picked up due to the optical path difference of
the two deflected trajectories \textit{exactly cancels} (is \textit{opposite
in sign} from) the \textquotedblleft Aharonov-Bohm phase\textquotedblright%
\ picked up by the same trajectories due to the flux that they enclose). This
opposite sign seems to have been rather unnoticed$\boldsymbol{:}$ In Feynman's
Fig.15-8 \cite{Feynman}, or in Felsager's Fig.2.16 \cite{Felsager}, the
classical trajectories are deflected after passing through a strip of a
magnetic field placed on the right of a double-slit apparatus. Both authors
determine the semiclassical phase picked up by the deflected trajectories and
find it consistent with the Aharonov-Bohm phase. One can see on closer
inspection, however, that the two phases actually \textit{have opposite signs
}(see our own Fig.2 and the discussion that follows below, where this is
proved in detail). Similarly, in the very recent review of Batelaan \&
Tonomura\cite{Batelaan}, their Fig.2 shows wavefronts associated to the
deflected classical trajectories where it is stated that \textquotedblleft the
phase shift calculated in terms of the Lorentz force is the same as that
predicted by the Aharonov-Bohm effect in terms of the vector
potential\textquotedblright. Once more, however, it turns out that the sign of
the classical phase-difference is really opposite to the sign of the
Aharonov-Bohm phase (see proof below). The phases are not equal as stated by
the authors. And it turns out that even \textquotedblleft electric
analogs\textquotedblright\ of the above cases also demonstrate this
opposite-sign relationship (see proof further below). All the above examples
can be viewed as a manifestation of the cancellations that have been found in
the present work for \textit{general} (even completely spread-out) quantum
states (but in those examples they are just special cases for narrow
wavepacket-states in classical motion).

Let us give a brief elementary proof of the above claimed opposite
sign-relationships$\boldsymbol{:}$ Indeed, in our Fig.2, the \textquotedblleft
Aharonov-Bohm phase\textquotedblright\ due to the flux enclosed between the
two classical trajectories (of a particle of charge $q$) is
\begin{equation}
\Delta\varphi^{AB}=2\pi\frac{q}{e}\frac{\Phi}{\Phi_{0}}, \label{ABphase}%
\end{equation}
with $\Phi_{0}=\frac{hc}{e}$ the flux quantum, and $\Phi\thickapprox BWd$ the
enclosed flux between the two trajectories (for small trajectory-deflections),
with the deflection originating from the presence of the magnetic strip $B$
and the associated Lorentz forces. On the other hand, the semiclassical phase
difference between the same 2 classical trajectories is $\Delta\varphi
^{semi}=\frac{2\pi}{\lambda}\Delta l$, with $\lambda=\frac{h}{mv}$ being the
de Broglie wavelength (and $v$ being the speed of the particle, taken almost
constant (as usually done) due to the small deflections), and with $\Delta l$
being $\Delta l\thickapprox d\sin\theta\thickapprox d\frac{x_{c}}{L}$ ($x_{c}$
being the (displaced) position of the central fringe on the screen). We have
therefore
\begin{equation}
\Delta\varphi^{semi}=\frac{2\pi}{\lambda}d\frac{x_{c}}{L}. \label{Semiphase}%
\end{equation}
Now, the Lorentz force (exerted only during the passage through the thin
magnetic strip, hence only during a time interval $\Delta t=\frac{W}{v}$) has
a component parallel to the screen (let us call it $x$-component) that is
given by
\begin{equation}
F_{x}=\frac{q}{c}(\boldsymbol{v}\times\boldsymbol{B})_{x}=-\frac{q}%
{c}vB=-\frac{BWq}{c\frac{W}{v}}=-\frac{BWq}{c\Delta t} \label{Fx}%
\end{equation}
which shows that there is a change of kinematic momentum (parallel to the
screen) equal to $-\frac{BWq}{c},$ or, equivalently, a change of parallel speed%

\begin{equation}
\Delta v_{x}=-\frac{BWq}{mc} \label{Dvx}%
\end{equation}
which is the speed of the central fringe's motion (i.e. its displacement over
time along the screen). Although this has been caused by the presence of the
thin deflecting magnetic strip, this displacement is occuring uniformly during
a time interval $t=\frac{L}{v},$ and this time interval must satisfy%

\begin{equation}
\Delta v_{x}=\frac{x_{c}}{t} \label{Dvx2}%
\end{equation}
(as, for small displacements, the wavepackets travel most of the time in
uniform motion, i.e. $\Delta t<<t$). We therefore have that the central fringe
displacement must be $x_{c}=\Delta v_{x}t=-\frac{BWq}{cm}\frac{L}{v},$ and
noting that $mv=\frac{h}{\lambda}$, we finally have%

\begin{equation}
x_{c}=-\frac{BWqL\lambda}{hc}. \label{xc}%
\end{equation}
By susbstituting (\ref{xc}) into (\ref{Semiphase}), the lengths $L$ and
$\lambda$ cancel out, and we finally have $\Delta\varphi^{semi}=-2\pi\frac
{q}{e}\frac{BWd}{\frac{hc}{e}},$ which with $\frac{hc}{e}=\Phi_{0}$ the flux
quantum, and $BWd\thickapprox\Phi$ the enclosed flux (always for small
trajectory-deflections) gives (through comparison with (\ref{ABphase})) our
final proof that%

\begin{equation}
\Delta\varphi^{semi}=-2\pi\frac{q}{e}\frac{\Phi}{\Phi_{0}}=-\Delta\varphi
^{AB}. \label{proof}%
\end{equation}
[It should be noted that such an opposite sign relation actually tells us that
the \textit{total} (semiclassical + AB)\textit{ }phase\textit{ difference} is
zero at the new position of the central (bright) fringe (after it has been
displaced due to the trajectories' deflection), and in this sense the above
minus sign should be rather expected.]

\bigskip%

\begin{figure}[ptb]%
\centering
\includegraphics[
height=4.7651in,
width=6.365in
]%
{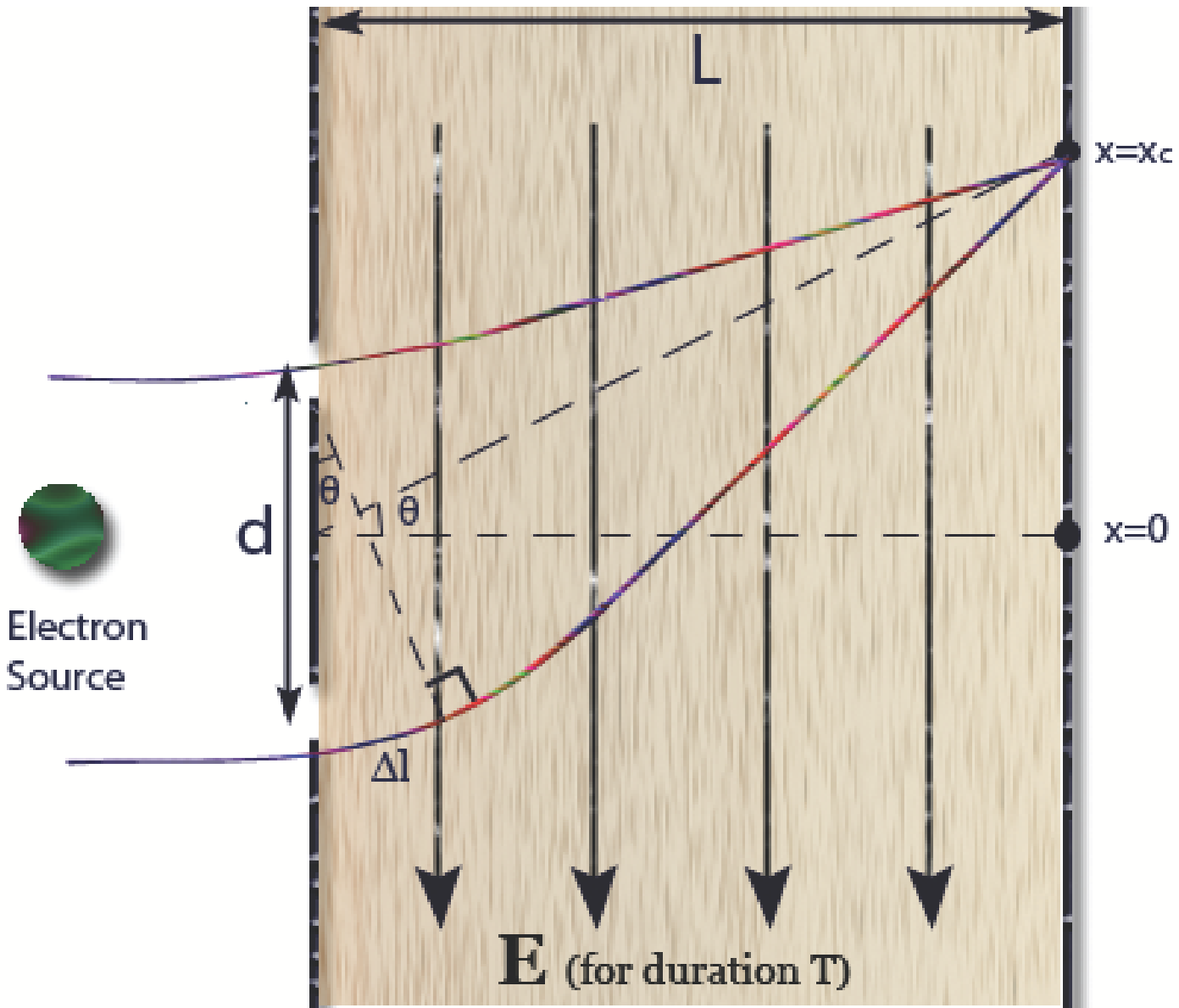}%
\end{figure}

\bigskip

The \textquotedblleft electric analog\textquotedblright\ of the above exercise
is also outlined below, now with a homogeneous electric field (pointing
downwards everywhere in space, but switched on for only a finite duration $T$)
on the right of a double-slit apparatus (see our Fig.3)$\boldsymbol{:}$ In
this case the electric Lorentz force $qE$ is exerted on the trajectories only
during the small time interval $\Delta t=T,$ which we take to be much shorter
($T<<t$) than the time of travel $t=\frac{L}{v}$ (we now have a thin electric
strip in \textit{time} rather than the thin magnetic strip in space that we
had earlier). The electric type of \textquotedblleft Aharonov-Bohm
phase\textquotedblright\ is now%

\begin{equation}
\Delta\varphi^{AB}=-2\pi\frac{q}{e}\frac{cT\Delta V}{\Phi_{0}}, \label{ABele}%
\end{equation}
with $\Delta V$ being the electric potential difference between the two
trajectories, hence $\Delta V\thickapprox Ed$ (again for small
trajectory-deflections). On the other hand, the semiclassical phase difference
between the two trajectories is again given by (\ref{Semiphase}), but the
position $x_{c}$ of the central fringe must now be determined by the electric
field force $qE\boldsymbol{:}$ The change of kinematic momentum (always
parallel to the screen) is now $qET$, hence the analog of (\ref{Dvx}) is now%

\begin{equation}
\Delta v_{x}=\frac{qET}{m} \label{Dvxnew}%
\end{equation}
which if combined with (\ref{Dvx2}) (that is obviously valid in this case as
well, again for small deflections, due to the $\Delta t=T<<t$), and always
with $t=\frac{L}{v}$, gives that the central fringe displacement must be
$x_{c}=\Delta v_{x}t=\frac{qET}{m}\frac{L}{v}$, and using again $mv=\frac
{h}{\lambda}$, we finally have the following analog of (\ref{xc})%

\begin{equation}
x_{c}=\frac{qETL\lambda}{h}. \label{xcnew}%
\end{equation}
By substituting (\ref{xcnew}) into (\ref{Semiphase}), the lengths $L$ and
$\lambda$ again cancel out, and we finally have $\Delta\varphi^{semi}=2\pi
d\frac{qETL\lambda}{h}=2\pi\frac{q}{e}\frac{EdcT}{\frac{hc}{e}}$, which with
$\frac{hc}{e}=\Phi_{0}$ the flux quantum, and through comparison with
(\ref{ABele}) leads once again to our final proof that%

\begin{equation}
\Delta\varphi^{semi}=-\Delta\varphi^{AB}. \label{proofnew}%
\end{equation}
We note therefore that even in the electric case, the semiclassical phase
difference (between two trajectories) picked up due to the Lorentz force
(exerted on them) is once again opposite to the electric \textquotedblleft
Aharonov-Bohm phase\textquotedblright\ phase picked up by the same
trajectories (due to the electric flux that they enclose).

We should point out once again, however, that although the above elementary
considerations apply to semiclassical motion of narrow wavepackets, in this
paper we have given \textit{a more general understanding of the above opposite
sign-relationships} that applies to general (even completely delocalized)
states, and that originates from what could be called \textquotedblleft
generalized Werner \& Brill cancellations\textquotedblright\ (the
cancellations that come out of our new solutions, in cases more general than
those of Werner \& Brill, including both magnetic and electric fields, both
static and $t$-dependent situations, and both semiclassical and spread-out
quantum states).

In a slightly different vein, the cancellations that we found above give an
explanation of why certain classical arguments (invoking the past
$t$-dependent history of an experimental setup) seem to be successful in
giving at the end an explanation of Aharonov-Bohm effects (namely a phase
consistent with that of a static Aharonov-Bohm configuration). However, there
is again an opposite sign that seems to have been largely unnoticed in such
arguments as well (i.e. in Silverman\cite{Silverman}, where in his eq.(1.34)
there should be an extra minus sign). Our above observation essentially
describes the fact that, \textit{if we had actually used} a $t$-dependent
magnetic flux (with its final value being the actual value of our static
flux), then the induced electric field (viewed now as a nonlocal term of the
present work) would have cancelled the static Aharonov-Bohm phase. Of course
now, this $t$-dependent experimental set up has not been used (the flux is
static) and we obtain the usual magnetic Aharonov-Bohm phase, but the above
argument (of a \textquotedblleft potential experiment\textquotedblright\ that
\textit{could have been carried out}) takes the \textquotedblleft
mystery\textquotedblright\ away of why such history-based arguments generally
work $-$ although \textit{they have to be corrected with a sign}. The above
also gives a rather natural account of the \textquotedblleft dynamical
nonlocality\textquotedblright\ character\cite{Popescu} attributed to the
various Aharonov-Bohm phenomena (magnetic, electric or combined), although $-$
in the present work $-$ this dynamical quantum nonlocality seems to
simultaneously respect Causality, as will be seen in the next Section. This is
a rather pleasing characteristic of this theory that, as far as we are aware,
has no parallel in other formulations.

\section{Full (x,y,t)-case}

\bigskip Finally, let us look at the most general spatially-two-dimensional
and time-dependent case. This combines effects of (perpendicular) magnetic
fields (which, if present only in physically-inaccessible regions, can have
Aharonov-Bohm consequences) with the temporal nonlocalities of electric fields
(parallel to the plane) found in previous Sections. By working again in
Cartesian spatial coordinates, we now have to deal with the full system of PDEs%

\begin{equation}
\frac{\partial\Lambda(x,y,t)}{\partial x}=A_{x}(x,y,t),\qquad\frac
{\partial\Lambda(x,y,t)}{\partial y}=A_{y}(x,y,t),\qquad-\frac{1}{c}%
\frac{\partial\Lambda(x,y,t)}{\partial t}=\phi\left(  x,y,t\right)  .
\label{FullSystem}%
\end{equation}
This exercise is considerably longer than the previous ones but important to
solve, in order to see in what manner the solutions of this system are able to
\textit{combine} the spatial and temporal nonlocal effects found above. There
are now 3!=6 alternative integration routes to follow for solving this system
(and, in addition to this, the results in intermediate steps tend to
proliferate). The corresponding (rather long) procedure for solving the system
(\ref{FullSystem}) is described in detail in \cite{jphysa}, and 2 out of the
12 solutions that can be derived turn out to be the most crucial for the
discussion that will follow. First, by following steps similar to the ones of
Section IV, the following temporal generalization of (\ref{Lambda(x,y)4}) is obtained%

\[
\Lambda(x,y,t)=\Lambda(x_{0},y_{0},t)+\int_{x_{0}}^{x}A_{x}(x^{\prime}%
,y_{0},t)dx^{\prime}+\int_{y_{0}}^{y}A_{y}(x,y^{\prime},t)dy^{\prime}+
\]

\begin{equation}
+\left\{  {\Huge -}%
{\displaystyle\int\limits_{x_{0}}^{x}}
dx^{\prime}%
{\displaystyle\int\limits_{y_{0}}^{y}}
dy^{\prime}B_{z}(x^{\prime},y^{\prime},t)+G(y,t)\right\}  +f(x_{0},t)
\label{Lambda(x,y,t)213}%
\end{equation}

\[
with\text{ \ }G(y,t)\text{ \ }such\text{ }that\text{ \ \ }\left\{  {\Huge -}%
{\displaystyle\int\limits_{x_{0}}^{x}}
dx^{\prime}%
{\displaystyle\int\limits_{y_{0}}^{y}}
dy^{\prime}B_{z}(x^{\prime},y^{\prime},t)+G(y,t)\right\}  \boldsymbol{:}\text{
\ is }\mathsf{independent\ of\ }\ y,
\]
and from this point on, the third equation of the system (\ref{FullSystem}) is
getting involved to determine the nontrivial effect of scalar potentials on
$G(y,t)\boldsymbol{.}$ Indeed, by combining it with (\ref{Lambda(x,y,t)213})
there results a wealth of patterns, one of them leading finally to our first
solution, namely%

\[
\Lambda(x,y,t)=\Lambda(x_{0},y_{0},t_{0})+\int_{x_{0}}^{x}A_{x}(x^{\prime
},y_{0},t)dx^{\prime}+\int_{y_{0}}^{y}A_{y}(x,y^{\prime},t)dy^{\prime}-%
{\displaystyle\int\limits_{x_{0}}^{x}}
dx^{\prime}%
{\displaystyle\int\limits_{y_{0}}^{y}}
dy^{\prime}B_{z}(x^{\prime},y^{\prime},t)+G(y,t_{0})-
\]

\begin{equation}
-c%
{\displaystyle\int\limits_{t_{0}}^{t}}
\phi(x_{0},y_{0},t^{\prime})dt^{\prime}+c%
{\displaystyle\int\limits_{t_{0}}^{t}}
dt^{\prime}%
{\displaystyle\int\limits_{x_{0}}^{x}}
dx^{\prime}E_{x}(x^{\prime},y,t^{\prime})+c%
{\displaystyle\int\limits_{t_{0}}^{t}}
dt^{\prime}%
{\displaystyle\int\limits_{y_{0}}^{y}}
dy^{\prime}E_{y}(x_{0},y^{\prime},t^{\prime})+F(x,y)+f(x_{0},t_{0})
\label{LambdaFull1}%
\end{equation}
with the functions $G(y,t_{0})$ \ and $\ F(x,y)$ \ to be chosen in such a way
as to satisfy the following 3 independent conditions$\boldsymbol{:}$%

\begin{equation}
\left\{  G(y,t_{0})-%
{\displaystyle\int\limits_{x_{0}}^{x}}
dx^{\prime}%
{\displaystyle\int\limits_{y_{0}}^{y}}
dy^{\prime}B_{z}(x^{\prime},y^{\prime},t_{0})\right\}  \boldsymbol{:}%
\ is\text{ \ }\mathsf{independent\ of\ }\ y, \label{Gcondition}%
\end{equation}
which is of course a special case of the condition on $G(y,t)$ above (see
after (\ref{Lambda(x,y,t)213})) applied at $t=t_{0}$, and the other 2 turn out
to be of the form%

\begin{equation}
\left\{  F(x,y)+c%
{\displaystyle\int\limits_{t_{0}}^{t}}
dt^{\prime}%
{\displaystyle\int\limits_{x_{0}}^{x}}
dx^{\prime}E_{x}(x^{\prime},y,t^{\prime})\right\}  \boldsymbol{:}\ is\text{
\ }\mathsf{independent\ of\ }\ x, \label{F(x,y)condition1}%
\end{equation}

\begin{equation}
\left\{  F(x,y)+c%
{\displaystyle\int\limits_{t_{0}}^{t}}
dt^{\prime}%
{\displaystyle\int\limits_{y_{0}}^{y}}
dy^{\prime}E_{y}(x,y^{\prime},t^{\prime})\right\}  \boldsymbol{:}\ is\text{
\ }\mathsf{independent\ of\ }\ y. \label{F(x,y)condition2}%
\end{equation}
(It is probably important to note that for the above results the Faraday's law
is crucial\cite{jphysa}). As for the constant quantity\ $f(x_{0},t_{0})$
appearing in (\ref{LambdaFull1}), it again describes possible effects of
multiple-connectivity at the instant $t_{0}$ (which are absent for
simple-connected spacetimes, but will be crucial in the discussion of the van
Kampen thought-experiment to be discussed later below).

\bigskip Eq. (\ref{LambdaFull1}) is our first solution. It is now crucial to
note that an alternative form of solution (with the functions $G$ and $F$
satisfying the \textit{same} conditions as above) can be derived, and it turns
out to be%

\[
\Lambda(x,y,t)=\Lambda(x_{0},y_{0},t_{0})+\int_{x_{0}}^{x}A_{x}(x^{\prime
},y_{0},t)dx^{\prime}+\int_{y_{0}}^{y}A_{y}(x,y^{\prime},t)dy^{\prime}-%
{\displaystyle\int\limits_{x_{0}}^{x}}
dx^{\prime}%
{\displaystyle\int\limits_{y_{0}}^{y}}
dy^{\prime}B_{z}(x^{\prime},y^{\prime},t_{0})+G(y,t_{0})-
\]

\begin{equation}
-c%
{\displaystyle\int\limits_{t_{0}}^{t}}
\phi(x_{0},y_{0},t^{\prime})dt^{\prime}+c%
{\displaystyle\int\limits_{t_{0}}^{t}}
dt^{\prime}%
{\displaystyle\int\limits_{x_{0}}^{x}}
dx^{\prime}E_{x}(x^{\prime},y_{0},t^{\prime})+c%
{\displaystyle\int\limits_{t_{0}}^{t}}
dt^{\prime}%
{\displaystyle\int\limits_{y_{0}}^{y}}
dy^{\prime}E_{y}(x,y^{\prime},t^{\prime})+F(x,y)+f(x_{0},t_{0}).
\label{LambdaFull2}%
\end{equation}
In this alternative solution we note that, in comparison with
(\ref{LambdaFull1}), the line-integrals of $\ \boldsymbol{E}$ \ have changed
to the \textit{other} alternative \textquotedblleft path\textquotedblright%
\ (note the difference in the placement of the coordinates of the initial
point $(x_{0},y_{0})$ in the arguments of $E_{x}$ and $E_{y}$ compared to
(\ref{LambdaFull1})) and they happen to have the same sense as the
$\boldsymbol{A}$-integrals, while simultaneously the magnetic flux difference
shows up with its value at the initial time $t_{0}$ rather than at $t$. This
alternative form will be shown to be useful in cases where we want to directly
compare physical situations in the present (at time $t$) and in the past (at
time $t_{0}$), and the above noted change of sense of $\boldsymbol{E}%
$-integrals (compared to (\ref{LambdaFull1})) will be crucial in the
discussion that follows in the next Section. (It is also important here to
note that, in the form (\ref{LambdaFull2}), the electric fields have already
incorporated the effect of radiated $B_{z}$-fields in space (through the
Maxwell's equations, see \cite{jphysa}), and this is why at the end only the
$B_{z}$ at $t_{0}$ appears explicitly).

Once again the reader can directly verify that (\ref{LambdaFull1}) or
(\ref{LambdaFull2}) indeed satisfy the basic input system (\ref{FullSystem}).
(This verification is considerably more tedious than the earlier ones but
rather straightforward).

\bigskip But a last mathematical step remains$\boldsymbol{:}$ in order to
discuss the van Kampen case, namely an enclosed (and physically inaccessible)
magnetic flux (which however is \textit{time-dependent})\textit{, }it is
important to have the analogous forms through a reverse route of integrations
(see \cite{jphysa}), where at the end we will have the reverse
\textquotedblleft path\textquotedblright\ of $\boldsymbol{A}$-integrals (so
that by taking the \textit{difference} of the resulting solution and the above
solution (\ref{LambdaFull1}) (or (\ref{LambdaFull2})) will lead to the
\textit{closed} line integral of $\boldsymbol{A}$,\textbf{ }which will then be
immediately related to van Kampen's magnetic flux (at the instant $t$)). By
following then the reverse route, and by applying a similar strategy at every
intermediate step, we finally obtain the following solution (the spatially
\textquotedblleft dual\textquotedblright\ of (\ref{LambdaFull1})), namely%

\[
\Lambda(x,y,t)=\Lambda(x_{0},y_{0},t_{0})+\int_{x_{0}}^{x}A_{x}(x^{\prime
},y,t)dx^{\prime}+\int_{y_{0}}^{y}A_{y}(x_{0},y^{\prime},t)dy^{\prime}+%
{\displaystyle\int\limits_{x_{0}}^{x}}
dx^{\prime}%
{\displaystyle\int\limits_{y_{0}}^{y}}
dy^{\prime}B_{z}(x^{\prime},y^{\prime},t)+\hat{G}(x,t_{0})-
\]

\begin{equation}
-c%
{\displaystyle\int\limits_{t_{0}}^{t}}
\phi(x_{0},y_{0},t^{\prime})dt^{\prime}+c%
{\displaystyle\int\limits_{t_{0}}^{t}}
dt^{\prime}%
{\displaystyle\int\limits_{x_{0}}^{x}}
dx^{\prime}E_{x}(x^{\prime},y_{0},t^{\prime})+c%
{\displaystyle\int\limits_{t_{0}}^{t}}
dt^{\prime}%
{\displaystyle\int\limits_{y_{0}}^{y}}
dy^{\prime}E_{y}(x,y^{\prime},t^{\prime})+F(x,y)+\hat{h}(y_{0},t_{0})
\label{LambdaFull4}%
\end{equation}
with the functions $\hat{G}(x,t_{0})$ \ and $\ F(x,y)$ \ to be chosen in such
a way as to satisfy the following 3 independent conditions$\boldsymbol{:}$%

\begin{equation}
\left\{  \hat{G}(x,t_{0})+%
{\displaystyle\int\limits_{y_{0}}^{y}}
dy^{\prime}%
{\displaystyle\int\limits_{x_{0}}^{x}}
dx^{\prime}B_{z}(x^{\prime},y^{\prime},t_{0})\right\}  \boldsymbol{:}%
\ is\ \ \mathsf{independent\ of\ }\ x, \label{Gcondition3}%
\end{equation}

\bigskip%
\begin{equation}
\left\{  F(x,y)+c%
{\displaystyle\int\limits_{t_{0}}^{t}}
dt^{\prime}%
{\displaystyle\int\limits_{x_{0}}^{x}}
dx^{\prime}E_{x}(x^{\prime},y,t^{\prime})\right\}  \boldsymbol{:}%
\ is\ \ \mathsf{independent\ of\ }\ x, \label{F(x,y)condition3}%
\end{equation}

\bigskip%
\begin{equation}
\left\{  F(x,y)+c%
{\displaystyle\int\limits_{t_{0}}^{t}}
dt^{\prime}%
{\displaystyle\int\limits_{y_{0}}^{y}}
dy^{\prime}E_{y}(x,y^{\prime},t^{\prime})\right\}  \boldsymbol{:}%
\ is\ \ \mathsf{independent\ of\ }\ y, \label{F(x,y)condition4}%
\end{equation}
where again for the above results the Faraday's law is crucial. The
corresponding analog of the alternative form (\ref{LambdaFull2}) (where
$B_{z}$ appears at $t_{0}$) is more important for the discussion that follows
and turns out to be%

\[
\Lambda(x,y,t)=\Lambda(x_{0},y_{0},t_{0})+\int_{x_{0}}^{x}A_{x}(x^{\prime
},y,t)dx^{\prime}+\int_{y_{0}}^{y}A_{y}(x_{0},y^{\prime},t)dy^{\prime}+%
{\displaystyle\int\limits_{x_{0}}^{x}}
dx^{\prime}%
{\displaystyle\int\limits_{y_{0}}^{y}}
dy^{\prime}B_{z}(x^{\prime},y^{\prime},t_{0})+\hat{G}(x,t_{0})-
\]

\begin{equation}
-c%
{\displaystyle\int\limits_{t_{0}}^{t}}
\phi(x_{0},y_{0},t^{\prime})dt^{\prime}+c%
{\displaystyle\int\limits_{t_{0}}^{t}}
dt^{\prime}%
{\displaystyle\int\limits_{x_{0}}^{x}}
dx^{\prime}E_{x}(x^{\prime},y,t^{\prime})+c%
{\displaystyle\int\limits_{t_{0}}^{t}}
dt^{\prime}%
{\displaystyle\int\limits_{y_{0}}^{y}}
dy^{\prime}E_{y}(x_{0},y^{\prime},t^{\prime})+F(x,y)+\hat{h}(y_{0},t_{0})
\label{LambdaFIN}%
\end{equation}
with $\hat{G}(x,t_{0})$ and $F(x,y)$ following the same 3 conditions above.
The constant term $\hat{h}(y_{0},t_{0})$ again describes possible
multiplicities at the instant $t_{0}\boldsymbol{;}$ it is absent for
simple-connected spacetimes, but will be crucial in the discussion of the van
Kampen thought-experiment.

In (\ref{LambdaFull4}) (and in (\ref{LambdaFIN})), note the \textquotedblleft
alternative paths\textquotedblright\ (compared to solution (\ref{LambdaFull1})
(and (\ref{LambdaFull2}))) of line integrals of $\boldsymbol{A}$'s (or of
$\boldsymbol{E}$'s). But the most crucial element for what follows is the need
to \textit{exclusively} use the forms (\ref{LambdaFull2}) and (\ref{LambdaFIN}%
) (where $B_{z}$ only appears at $t_{0}$), and the fact that, within each
solution, the sense of $\boldsymbol{A}$-integrals is the \textit{same} as the
sense of the $\boldsymbol{E}$-integrals. (This is \textit{not} true in the
other solutions where $B_{z}(..,t)$ appears, as the reader can directly see).
These facts will be crucial to the discussion that follows, which briefly
addresses the so called \textquotedblleft van Kampen paradox\textquotedblright.

\bigskip In \cite{vanKampen} van Kampen considered a magnetic Aharonov-Bohm
setup, but with an inaccessible magnetic flux that is $t$%
-dependent$\boldsymbol{:}$ he envisaged turning on the flux very late, or
equivalently, observing the interference of the two wavepackets on a distant
screen very early, earlier than the time it takes light to travel the distance
to the screen (i.e. $t<\frac{R}{c}$), hence using the (instantaneous nature of
the) Aharonov-Bohm phase to transmit information (on the presence of a
confined flux somewhere in space) \textit{superluminally}. Indeed, the
Aharonov-Bohm phase at any $t$ is determined by differences of $\frac{q}{\hbar
c}\Lambda(\mathbf{r},t)$ with $\Lambda(\mathbf{r},t)\sim\int_{\mathbf{r}_{0}%
}^{\mathbf{r}}\mathbf{A}(\mathbf{r}^{\prime},t)\boldsymbol{.}d\mathbf{r}%
^{\prime}$ (basically a special case of (\ref{wrong})). However, if we use,
instead, our results (\ref{LambdaFull2}) and (\ref{LambdaFIN}) above (that
contain the additional nonlocal terms), it turns out that, for a
spatially-confined magnetic flux $\Phi(t)$, the functions $G,$ $\hat{G}$ and
$F$ in the above solutions can then all be taken zero$\boldsymbol{:}$
\textbf{(i)} their conditions are all satisfied for a flux $\Phi(t)$ that is
not spatially-extended (hence, from the 2 conditions on $G$ and $\hat{G}$ (eq.
(\ref{Gcondition}) and (\ref{Gcondition3})) we obtain $G=\hat{G}=0$ since the
integrals in brackets are all independent of $x$ and $y$), and \textbf{(ii)}
for $t<$ $\frac{R}{c}$, the integrals of $E_{x}$ and $E_{y}$ in the
corresponding conditions (eq.(\ref{F(x,y)condition1}) and
(\ref{F(x,y)condition2})) are already independent of \textit{both} $x$ and $y$
(since $E_{x}(x,y,t^{\prime})=E_{y}(x,y,t^{\prime})=0$ for all $t^{\prime
}<t<\frac{R}{c}$, with $(x,y)$ the observation point [the essential point
being that at instant $t$, the $\boldsymbol{E}$-field has not yet reached the
spatial point $(x,y)$ of the screen, and therefore all integrations of $E_{x}$
and $E_{y}$ with respect to $x^{\prime}$ and $y^{\prime}$ will be contributing
only up to a light-cone (see Fig.4) and they will therefore give results that
are \textit{independent of the integration upper limits }$x$\textit{ and }$y$
$-$ basically a generalization of the striped cases that we saw earlier but
now to the case of 3 spatio-temporal variables (with now the spatial point
$(x,y)$ being outside the light-cone defined by $t$ (see Fig.4; in this figure
the initial spatial point $(x_{0},y_{0})$, taken for simplicity at $(0,0)$,
has been supposed to be in the area of the inaccessible flux $\Phi(t)$, so
that, for $\sqrt{(x-x_{0})^{2}+(y-y_{0})^{2}}=\sqrt{x^{2}+y^{2}}=R$, we have
indeed that $ct<R$, as written on the figure))]$\boldsymbol{;}$ we therefore
rigorously obtain $F=0$). Moreover, the Aharonov-Bohm multiplicities (at
$t_{0}$) lead to cancellation of the $B_{z}$-terms (always at $t_{0}$), with
the final result (after subtraction of the 2 solutions) being%

\[
\Delta\Lambda(x,y,t)=\int_{x_{0}}^{x}A_{x}(x^{\prime},y_{0},t)dx^{\prime}%
+\int_{y_{0}}^{y}A_{y}(x,y^{\prime},t)dy^{\prime}-\int_{x_{0}}^{x}%
A_{x}(x^{\prime},y,t)dx^{\prime}-\int_{y_{0}}^{y}A_{y}(x_{0},y^{\prime
},t)dy^{\prime}+
\]

\begin{equation}
+c%
{\displaystyle\int\limits_{t_{0}}^{t}}
dt^{\prime}\left\{
{\displaystyle\int\limits_{x_{0}}^{x}}
dx^{\prime}E_{x}(x^{\prime},y_{0},t^{\prime})+%
{\displaystyle\int\limits_{y_{0}}^{y}}
dy^{\prime}E_{y}(x,y^{\prime},t^{\prime})-%
{\displaystyle\int\limits_{x_{0}}^{x}}
dx^{\prime}E_{x}(x^{\prime},y,t^{\prime})-%
{\displaystyle\int\limits_{y_{0}}^{y}}
dy^{\prime}E_{y}(x_{0},y^{\prime},t^{\prime})\right\}  . \label{DeltaLambda}%
\end{equation}

This can equivalently be written as%
\begin{equation}
\Delta\Lambda(x,y,t)=%
{\displaystyle\oint}
\mathbf{A}(\mathbf{r}^{\prime},t)\boldsymbol{.}d\mathbf{r}^{\prime}+c%
{\displaystyle\int\limits_{t_{0}}^{t}}
dt^{\prime}%
{\displaystyle\oint}
\mathbf{E}(\mathbf{r}^{\prime},t^{\prime})\boldsymbol{.}d\mathbf{r}^{\prime}
\label{DeltaLambdaBrief}%
\end{equation}
which, with $%
{\displaystyle\oint}
\mathbf{A}(\mathbf{r}^{\prime},t)\boldsymbol{.}d\mathbf{r}^{\prime}=\Phi(t)$
\ the instantaneous enclosed magnetic flux and with the help of Faraday's law
$%
{\displaystyle\oint}
\mathbf{E}(\mathbf{r}^{\prime},t^{\prime})\boldsymbol{.}d\mathbf{r}^{\prime
}=-\frac{1}{c}\frac{d\Phi(t^{\prime})}{dt^{\prime}},$ \ gives%

\begin{equation}
\Delta\Lambda(x,y,t)=\Phi(t)-{\huge (}\Phi(t)-\Phi(t_{0}){\huge )}=\Phi
(t_{0}). \label{DeltaLambdaFinal}%
\end{equation}
Although $\Delta\Lambda$ is generally $t$-dependent, we obtain the intuitive
(causal) result that, for $t<\frac{L}{c}$ (i.e. if the physical information
has not yet reached the screen), the phase-difference turns out to be
$t$-independent, and leads to the magnetic Aharonov-Bohm\ phase that we
\textit{would} observe at $t_{0}$. \textit{The new nonlocal terms have
conspired in such a way as to exactly cancel the Causality-violating
Aharonov-Bohm phase} (that would be proportional to the instantaneous
$\Phi(t)$). This gives an honest resolution of the van Kampen
\textquotedblleft paradox\textquotedblright\ within a canonical formulation,
without using any vague electric Aharonov-Bohm argument (as there is no
multiple-connectivity in the $(x,t)$-plane\cite{Iddings}). An additional
physical element is that, for the above cancellation, it is not only the
$E$-fields but also the $t$-propagation of the $B_{z}$-fields (the full
\textquotedblleft radiation field\textquotedblright) that plays a
role\cite{jphysa}.

\bigskip%

\begin{figure}[ptb]%
\centering
\includegraphics[
height=4.7651in,
width=6.365in
]%
{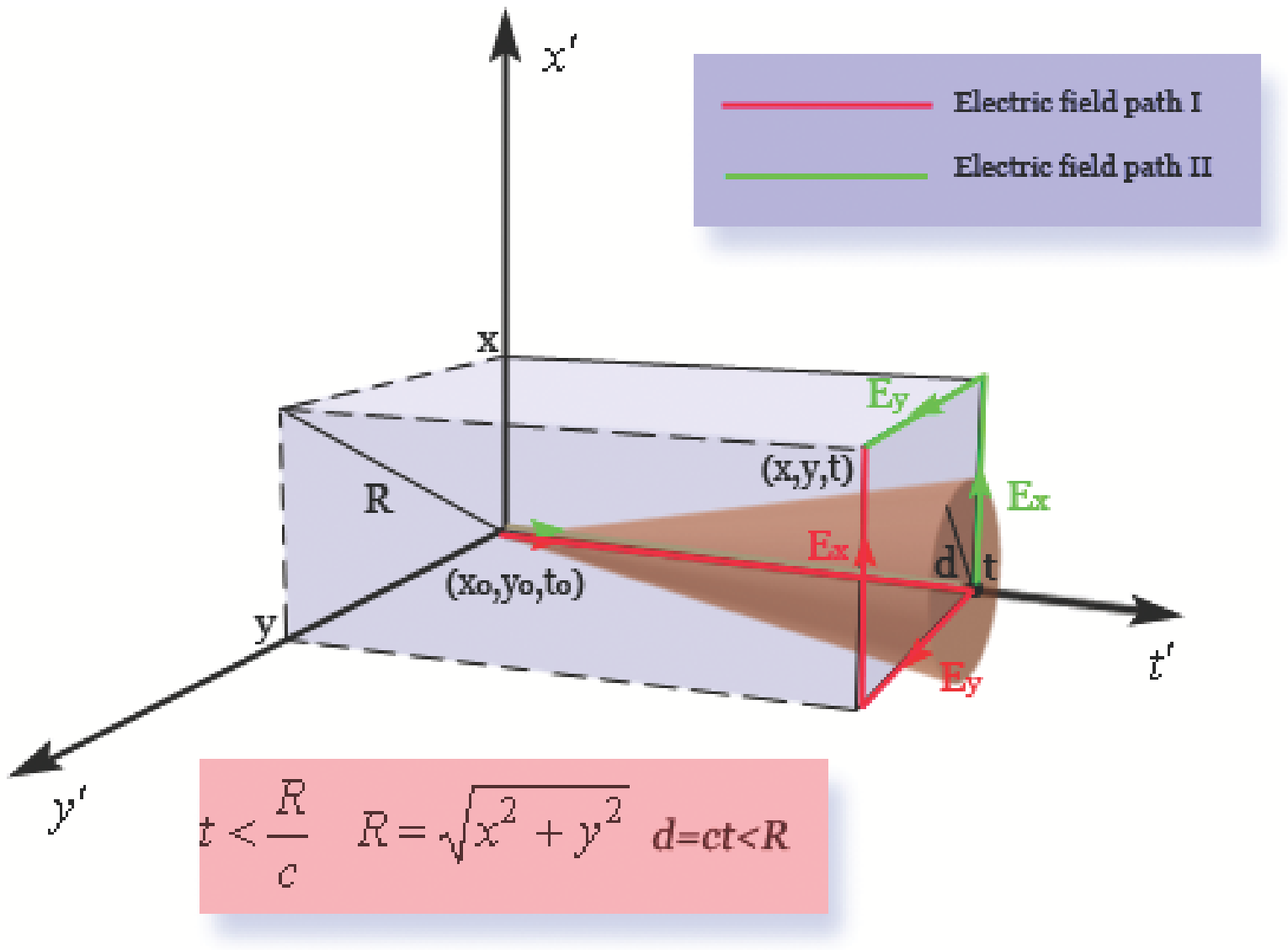}%
\end{figure}

\bigskip

Use of other 10 solutions that rigorously come out from the basic system of
PDEs can also address bound-state analogs (i.e. in $t$-dependent magnetic
flux-driven 1-D nanorings, such as in \cite{LuanTang}) or even
\textquotedblleft electric\textquotedblright\ analogs of the above van Kampen
case$\boldsymbol{:}$ In Peshkin's review\cite{Peshkin} on the electric
Aharonov-Bohm effect,$\boldsymbol{\ }$the author correctly states
\textquotedblleft One cannot wait for the electron to pass and only later
switch on the field to cause a physical effect\textquotedblright. Although
Peshkin uses his eq.(B.5) and (B.6) (based on the incorrect (\ref{wrong})), he
carefully states that it is not the full solution; actually, if we view it as
an \textit{ansatz}, then it is understandable why he needs to enforce a
\textit{condition} (his eq.(B.8), and later (B.9)) on the electric field
outside the cages (in order for certain (annoying) terms (resulting from a
minimal substitution due to the incorrect ansatz) to vanish and for (B.5) to
be a solution). But then he notes that the extra condition cannot always be
satisfied (hence (B.5) is not really the solution for all times), drawing from
this the above qualitatively correct conclusion on Causality. As it turns out,
our treatment gives exactly what Peshkin describes in words (with the total
\textquotedblleft radiation field\textquotedblright\ outside the cages being
once again crucial in recovering Causality), in a similar way as in the case
presented above in this Section for the usual (magnetic) version of the van
Kampen experiment. In this \textquotedblleft electric analog\textquotedblright%
\ that we are discussing now, the causally-offending part of the electric
Aharonov-Bohm phase difference will be cancelled by a magnetic type of phase,
that originates from the magnetic field that is associated with the
$t$-dependence of the electric field $\boldsymbol{E}$ outside the cages. We
conclude that our (exact) results accomplish precisely what Peshkin has in
mind in his discussion (on Causality), but in a direct and fully quantitative
manner, and with \textit{no ansatz} based on an incorrect form.

\bigskip

\section{\bigskip Discussion}

Returning for the moment to only one of the many misconceptions briefly
pointed out in this paper, we should emphasize further that improper uses of
simple Dirac phases in the literature are not rare or marginal$\boldsymbol{:}$
even in Feynman\cite{Feynman} it is stated that the simple phase factor
$\int^{x}\boldsymbol{A}\cdot d\mathbf{r}^{\prime}-c\int^{t}\phi dt^{\prime}$
is valid even for dynamic fields; this is also explicitly stated in
Erlichson's review\cite{Erlichson} $-$ Silverman\cite{Silverman} being the
only report that we are aware of with a careful wording about (\ref{wrong})
being only restrictedly valid (for $t$-independent $\mathbf{A}$ and
$\mathbf{r}$-independent $\phi$), although even there the nonlocal terms have
been missed.

With respect to the presence of \textit{fields} in the \textit{phases} of
quantum mechanical wavefunctions that we find, it should be stressed that at
the level of the basic Lagrangian $L(\mathbf{r},\mathbf{v},t)=\frac{1}%
{2}m\mathbf{v}^{2}+\frac{q}{c}\mathbf{v}\boldsymbol{.}\mathbf{A}%
(\mathbf{r},t)-q\phi(\mathbf{r},t)$ \ there are no fields present, and the
view holds in the literature\cite{BrownHome} that electric fields or magnetic
fields cannot contribute directly to the phase. This view originates from the
path-integral treatments widely used (where the Lagrangian determines directly
the phases of Propagators), but, nevertheless, our canonical treatment shows
that fields \textit{do} contribute nonlocally, and they are actually crucial
in recovering Relativistic Causality. Moreover, path-integral
discussions\cite{Troudet} of the van Kampen case use wave (retarded)-solutions
for $\mathbf{A}$ (hence in the Lorenz gauge) and are incomplete; our results
take advantage of the retardation of \textit{fields} $\mathbf{E}$ and
$\mathbf{B}$\textbf{ }(true in \textit{any} gauge), and \textit{not }of
potentials. In addition, Troudet\cite{Troudet} correctly states that his
path-integral treatment is good for not highly-delocalized states in space,
and that in case of delocalization the proper treatment \textquotedblleft
would be much more complicated, and would require a much more complete
analysis\textquotedblright. It is fair to state that such a complete analysis
has actually been provided in the present work. It should be added that the
van Kampen \textquotedblleft paradox\textquotedblright\ seems to be still
thought of as \textquotedblleft remarkable\textquotedblright%
\ \cite{Compendium}. The present work has provided a natural and general
\textit{resolution}, and most importantly, through nonlocal (and
Relativistically causal) propagation of wavefunction-phases.

Finally, in trying to explore an even broader significance of the new
solutions, one may wonder about possible consequences of the nonlocal terms if
these are included in more general physical models that have a gauge structure
(in Condensed Matter or High Energy Physics). It is also worth mentioning that
if one follows the same \textquotedblleft unconventional\textquotedblright%
\ method (of solution of PDEs) with the Maxwell's equations for the electric
and magnetic fields (rather than with the PDEs of eq.(\ref{Basic11}) for the
potentials that give $\Lambda$), the corresponding nonlocal terms can be
derived in a similar manner, and one can then see that these nonlocal terms
essentially demonstrate the causal propagation of the radiation electric and
magnetic fields outside physically inaccessible confined sources (i.e.
solenoids or electric cages). Although this is of course widely known at the
level of classical fields, a major conclusion that can be drawn from the
present work (at the level of gauge transformations) is that \textit{a
corresponding Causality also exists at the level of quantum mechanical phases}
as well, and this is enforced by the nonlocal terms in $t$-dependent cases. It
strongly indicates that the nonlocal terms found in this work at the level of
quantum mechanical phases reflect a causal propagation of wavefunction phases
\textbf{in the Schr\"{o}dinger Picture} (at least one part of them $-$ the one
containing the fields $-$ which competes with the Aharonov-Bohm types of
phases containing the potentials). This is an entirely new concept (given the
\textit{local nature} but also the \textit{nonrelativistic character} of the
Schr\"{o}dinger equation) and deserves to be further explored. It would indeed
be worth investigating possible applications of the above results (of nonlocal
phases of wavefunctions, solutions of the local Schr\"{o}dinger equation) to
$t$-dependent single- \textit{vs} double-slit experiments recently discussed
by the group of Aharonov\cite{Tollaksen} who use a completely different
method, with modular variables in the Heisenberg picture (presented as the
sole method appropriate for problems of this type). One should also note
recent work\cite{He}, that rightly emphasizes that Physics cannot currently
predict how we dynamically go from the single-slit diffraction pattern to the
double-slit diffraction pattern (whether it is in a gradual and causal manner
or not) and where a relevant experiment is proposed to decide on (address)
exactly this largely unknown issue. Application of our nonlocal terms to such
questions in analogous experiments (i.e. by introducing (finite) scalar
potentials on slits in a $t$-dependent way) provides a completely new
formulation for addressing causal issues of this type, and is worth of further
investigation. Furthermore, it is worth noting that, if $E$'s were substituted
by gravitational fields and $B$'s by Coriolis force fields arising in
non-inertial (rotating) frames of reference, the above nonlocalities (and
their apparent causal nature) could possibly have an interesting story to tell
about quantum mechanical phase behavior in a Relativistic/Gravitational
framework. Finally, \textit{SU(2)} generalizations would be an obviously
interesting extension of the above \textit{U(1)} theory, and such
generalizations are rather formally direct and not difficult to
make$\boldsymbol{;}$ an immediate physically interesting question would then
be whether the new nonlocal terms might have a nontrivial impact on i.e.
spin-$\frac{1}{2}$-states, since these terms would act asymmetrically on
opposite spins (the nonlocal $B_{z}$-terms being relevant for Zeeman
interactions, and the nonlocal $E$-terms possibly having a role if the above
results were applied i.e. to Condensed Matter systems with strong spin-orbit
coupling\cite{HasanKane}).

\bigskip

\section{\bigskip Conclusions}

We conclude that a nonlocal and causal behavior exists at the level of quantum
mechanical phases, even for solutions of the nonrelativistic (and local)
Schr\"{o}dinger equation and this behavior is enforced by the nonlocal terms
derived in the present work (through the well-known causal behavior of
fields). Our (exact and analytical) results accomplish precisely what Peshkin
has in mind in his discussion (on Causality) of the electric Aharonov-Bohm
effect, but in a direct and fully quantitative manner, and with \textit{no
ansatz} based on an incorrect form. Another pleasing characteristic of our
results that, as far as we are aware, has no parallel in other formulations,
is that they give a rather natural account of the \textquotedblleft dynamical
nonlocality\textquotedblright\ character\cite{Popescu} attributed to the
various Aharonov-Bohm phenomena (magnetic, electric or combined), although $-$
in the present work $-$ this dynamical quantum nonlocality seems to
simultaneously respect Causality in a \textquotedblleft deterministic
way\textquotedblright, i.e. without requiring the necessity of invoking the
Uncertainty Principle. The nonlocal terms found in this work at the level of
$\Lambda$ reflect a causal propagation of wavefunction-phases \textit{in the
Schr\"{o}dinger picture}, with possible immediate applications to
$t$-dependent slit-experiments recently discussed using the Heisenberg
picture\cite{Tollaksen}. Application of our nonlocal terms to such problems
(i.e. by introducing $t$-dependent scalar potentials on the slits) provides a
new and \textit{direct} formulation for addressing causal issues of such
$t$-dependent slit-systems. Finally, one cannot refrain from wondering about
the analogs of these new nonlocalities in many areas of Physics where
geometric or topological phases\cite{ShapereWilczek} appear as the central
quantities (these always being of the form of integrals of some effective (or
emergent) potentials (that are determined by the system, i.e. by band
structures, as in the rapidly expanding area of Topological
Insulators\cite{HasanKane})). A natural thought then arising from the present
work is that, if those emerging potentials are not of a type that would
correspond to zero fields (in the physically accessible regions), but describe
nonvanishing effective fields in (even remote) spacetime regions that are
accessible to the particles, then one would expect that the new nonlocalities
should be seriously taken into account $-$ these having certain dynamical
consequences that may have not received an entirely proper treatment in
earlier works. Given the popularity and importance of these areas in the whole
of Physics, issues such as the above would certainly deserve further study.

\section{Acknowledgements}

Graduate students Kyriakos Kyriakou and Georgios Konstantinou of the
University of Cyprus and Dr. Areg Ghazaryan of Yerevan State University are
acknowledged for having carefully reproduced all results. Georgios
Konstantinou is also gratefully acknowledged for having drawn all figures of
this article. Dr. Cleopatra Christoforou of the Department of Mathematics and
Statistics of the University of Cyprus is acknowledged for a discussion
concerning the mathematical method followed.

\textbf{Figure 1.} (Color online)$\boldsymbol{:}$ Examples of simple
field-configurations (in simple-connected regions), where the nonlocal terms
exist and are nontrivial, but can easily be determined$\boldsymbol{:}$ (a) a
striped case in 1+1 spacetime, where the electric flux enclosed in the
\textquotedblleft observation rectangle\textquotedblright\ is dependent on $t$
but independent of $x$; (b) a triangular distribution in 2-D space, where the
part of the magnetic flux inside the corresponding \textquotedblleft
observation rectangle\textquotedblright\ depends on \textit{both} $x$
\textit{and} $y$. The appropriate choices for the corresponding nonlocal
functions $g(x)$ and $\hat{g}(t)$ for case (a), or $g(x)$ and $h(y)$ for case
(b), are given in the text (Sections III and IV respectively).

\textbf{Figure 2.} (Color online)$\boldsymbol{:}$ The standard double-slit
apparatus with an additional strip of a perpendicular magnetic field $B$ of
width $W$ placed between the slit-region and the observation screen. In the
text we deal for simplicity with the case $W<<L,$ so that deflections (of the
semiclassical trajectories) due to the Lorentz force, shown here for a
negative charge $q$, are very small.

\textbf{Figure 3.} (Color online)$\boldsymbol{:}$ The analog of Fig.2 (again
for a negative $q$) but with an additional electric field parallel to the
observation screen that is turned on for a time interval $T$. In the text we
deal for simplicity with the case $T<<\frac{L}{v}$ (with $v=\frac{1}{m}%
\frac{h}{\lambda}$, $\lambda$ the de Broglie wavelength), so that deflections
(of the semiclassical trajectories) due to the electric force are again very
small. For both Fig.2 and 3, it is shown in the text that $\Delta
\varphi^{semiclassical}=-\Delta\varphi^{AB},$ hence we observe an extra minus
sign compared to what is usually reported in the literature.

\textbf{Figure 4.} (Color online)$\boldsymbol{:}$ The analog of paths of Fig.1
but now in 2+1 spacetime for the van Kampen thought-experiment, when the
instant of observation $t$ is so short that the physical information has not
yet reached the spatial point of observation $(x,y)$. The two solutions (that,
for wavepackets, have to be subtracted in order to give the phase difference
at $(x,y,t)$) are given in the text, and are here characterized through their
electric field $E$-line integral behavior$\boldsymbol{:}$ \textquotedblleft
electric field path (I)\textquotedblright\ (the red-arrow route) denotes
solution (\ref{LambdaFIN}), and \textquotedblleft electric field path
(II)\textquotedblright\ (the green-arrow route) denotes solution
(\ref{LambdaFull2}). Note that the strips of Fig.1(a) have now given their
place to a light-cone. At the point of observation (that lies outside this
light-cone) the Aharonov-Bohm phase difference has now become
\textquotedblleft causal\textquotedblright\ due to cancellations between the
two solutions (the two \textquotedblleft electric field
paths\textquotedblright\ above).

\end{document}